\documentclass[twocolumn,showpacs,preprintnumbers,amsmath,amssymb,prb,superscriptaddress]{revtex4}
\usepackage{mathrsfs}
\usepackage{amsmath}
\usepackage{amsfonts}
\usepackage{bm}
\usepackage{amssymb}
\usepackage{graphicx}
\usepackage{dcolumn}
\usepackage{txfonts}
\usepackage{color}

\begin{document}

\title{Optical effects of spin currents in semiconductors}
\author{Jing Wang}
\thanks{Current address: Department of Physics, Stanford University, Stanford, CA 94305-4045, USA}
\affiliation{State Key Laboratory of Low-Dimensional Quantum Physics, and Department of Physics, Tsinghua University, Beijing 100084, China}
\affiliation{Department of Physics, The Chinese University of Hong Kong, Shatin, N.T., Hong Kong, China}
\author{Sheng-Nan Ji}
\affiliation{State Key Laboratory of Low-Dimensional Quantum Physics, and Department of Physics, Tsinghua University, Beijing 100084, China}
\affiliation{Department of Physics, The Chinese University of Hong Kong, Shatin, N.T., Hong Kong, China}
\author{Bang-Fen Zhu}
\thanks{bfz@mail.tsinghua.edu.cn}
\affiliation{State Key Laboratory of Low-Dimensional Quantum Physics, and Department of Physics, Tsinghua University, Beijing 100084, China}
\affiliation{Institute of Advanced Study, Tsinghua University, Beijing 100084, China}
\author{Ren-Bao Liu}
\thanks{rbliu@phy.cuhk.edu.hk}
\affiliation{Department of Physics, The Chinese University of Hong Kong, Shatin, N.T., Hong Kong, China}

\begin{abstract}
A spin current has novel linear and second-order nonlinear optical effects due to its symmetry properties. With the symmetry analysis and the eight-band microscopic calculation we have systematically investigated the interaction between a spin current and a polarized light beam (or the "photon spin current") in direct-gap semiconductors.
This interaction is rooted in the intrinsic spin-orbit coupling in valence bands and does not rely on the Rashba or Dresselhaus effect.
The light-spin current interaction results in an optical birefringence effect
of the spin current. The symmetry analysis indicates that in a semiconductor with inversion symmetry,
the linear birefringence effect vanishes and only the circular birefringence effect exists. The circular birefringence effect is similar to the Faraday rotation in magneto-optics but involves no net magnetization nor breaking the time-reversal symmetry.  Moreover, a spin current can induce the second-order nonlinear optical processes due to the inversion-symmetry breaking. These findings form a basis of measuring a pure spin current where and when it flows with the standard optical spectroscopy, which may provide a toolbox to explore a wealth of physics connecting the spintronics and photonics.
\end{abstract}

\date{\today}


\maketitle

\section{Introduction}
A pure spin current consists of flows of opposite spins in opposite
directions with the same amplitude. It bears neither net charge
current nor net spin polarization. Spin currents are a key element in spintronics.\cite{wolf2001science,zutic2004rmp}

Detection of spin currents is important for characterization and applications in future spintronics technologies.\cite{wolf2001science,zutic2004rmp} While a polarized spin current may be detected by the conventional
Faraday/Kerr rotation spectroscopy\cite{kikkawa1999nature,stephens2004prl,crooker2005science}
or through ferromagnetic filters,\cite{lou2007natphys,appelbaum2007nature} a
pure spin current, without a direct electromagnetic induction, is
much less traceable. Still, pure spin currents have been detected in
a few pioneering experiments in which they were converted into
signals detectable by conventional techniques. For example, the spin-polarized electrons or
excitons accumulated at the sample edges where a spin current is
terminated may be detected by the Faraday/Kerr
rotation,\cite{kato2004science,stern2008natphys} polarized light
emission,\cite{wunderlich2005prl} and polarization-selective
absorption.\cite{stevens2003prl,zhao2006prl} Or the inverse spin Hall
effect\cite{dyakonov1971pla,hirsch1999prl,murakami2003science,sinova2004prl} can
be used to covert a spin current into charge/voltage signals for
electric measurement.\cite{valenzuela2006nature,ganichev2007prb,cui2007apl}
All such measurements, however, disturb the spin currents to some
extent and are indirect. We are motivated to find a non-destructive
way to directly measure a pure spin current.\cite{wang2008prl,wang2010prl}

A basic symmetry principle states that
whenever there is a current breaking the fundamental symmetries of a
system, an interaction may arise between the current and another
current of the same symmetry-breaking type so that the fundamental
symmetries are retained.\cite{coleman1969pr,callan1969pr} A classic example is
the Amp\`{e}re effect and the {\O}rsted effect where a charge current
is coupled to another charge current or a magnet. A straightforward analogue suggests that a pure spin
current may be coupled to another spin current. Such an
idea stimulated the proposal of direct measurement of a pure spin
current in a direct-gap semiconductor by a polarized light
beam.\cite{wang2008prl} A polarized light beam can be regarded as such
a ``photon spin current''\cite{Note_on_pure_current} by mapping the
photon polarization into a spin-1/2 in the Jones vector
representation\cite{jones_vector1941}
\begin{align}\label{jonesvector}
\cos\frac{\theta}{2}e^{i\phi/2}{\mathbf n}_+
+\sin\frac{\theta}{2}e^{-i\phi/2}{\mathbf n}_-\sim
\left|\theta,\phi\right\rangle,
\end{align}
where the right/left circular polarization ${\mathbf n}_{+/-}$
corresponds to the spin up/down state
$\left|\uparrow/\downarrow\right\rangle$ quantized along the light
propagation direction. The effective interaction between a pure spin
current and a polarized light causes a phase delay which depends on
the light polarization and wavevector. The observable result is a
circular birefringence effect which is similar to the Faraday
rotation but involves no net magnetization nor time-reversal
symmetry breaking. Since Faraday's disocvery in 1845,\cite{Faraday} the circular birefringence effect of spin currents is the first example of Faraday rotation without time-reversal symmetry breaking. We dub this effect as {\em spin current Faraday effect}.

At this point, we should mention a recent remarkable experiment realizing
the direct in-situ detection of a spin current through the Doppler effect
of a spin-wave.\cite{vlaminck2005science} In fact, the observed Doppler
effect and our predicted optical birefringence effect are
fundamentally related to each other. The former is the frequency
shift of the spin wave, while the latter is the phase shift of the
light accumulated by a frequency shift over a coupling time. The
frequency shift is measured in the near-field as in the experiment,
while the phase shift should be measured in the far field by light
polarization detection. Fundamentally, both are due to the effective
coupling between a pure spin current and another ``probe spin current'',
either a spin wave or a polarized light, mediated by virtual
excitations in the systems.

The effective light-spin current interaction is induced in a
semiconductor by virtual excitations of electron-hole pairs. The
specific form of the phenomenological coupling depends on the
microscopic mechanisms.\cite{wang2008prl} Since the light polarization
essentially couples only to the orbital motion of electrons, the
spin-orbit interaction is needed to establish the effective
coupling. As there is inherent spin-orbit coupling in the valence
bands due to the relativity effect, the Rashba or Dresselhaus effect
due to the spatial inversion asymmetries\cite{rashba1984,dresselhaus1955pr,ganichev2004prl} is
not a necessity, thus the system can bear the inversion symmetry.

The optical birefringence effect of spin currents\cite{wang2008prl} is usually very weak, because a tiny light wave vector $\mathbf{q}$ is involved in the coupling to the velocity $\mathbf{v}$ of the pure spin currents. However, if the velocity of spin currents couples to another optical field,
\[
\mathbf{q}\cdot\mathbf{v}\Rightarrow\mathbf{F}_2\cdot\mathbf{v}
\]
the coupling will be much enhanced. This means we can use the second optical field to drive the spins, which may result in the nonlinear optics of the pure spin current. In fact, such an analogy stimulated the prediction of the second-order nonlinear optical effects of pure spin currents,\cite{wang2010prl} which was soon verified by experiments.\cite{weraka2010natphys}

In Refs.~\onlinecite{wang2008prl} and ~\onlinecite{wang2010prl} we have sketched the main ideas based on
symmetry arguments and given the key expressions in a special model neglecting
the energy band anisotropy. In this paper, we will investigate in a more
comprehensive way the linear and second-order nonlinear optical effects of pure spin currents, including a systematic symmetry analysis of all relevant physical quantities, and
a detailed derivation for the effective Hamiltonian as well as the second-order nonlinear optical susceptibility.
The microscopic derivation confirms the qualitative results obtained by the
symmetry analysis. In particular, both the symmetry analysis and
the microscopic calculation lead to the conclusion that the linear birefringence effect (similar to
the Voigt effect in magneto-optics) always vanishes and only the
circular birefringence effect exists, and the energy band anisotropy
induces only a relatively small quantitative modification of
the results. The absence of the Voigt effect is fundamentally related to the lack of the $|0\rangle$ state in the physical spin of photons [not the pseudo-spin in Eq.~(\ref{jonesvector})]. The microscopic mechanism of both linear and second-order nonlinear effects can be understood in a unified physical picture.

In this paper, we assume that the host semiconductor system has the inversion
symmetry. We note that in compound semiconductors such as GaAs the
inversion symmetry is broken, which, though a small effect, is
critical to some schemes of spin current
generation.\cite{bhat2000prl,bhat2005prl} In our present scheme, however, the small inversion asymmetry in
compound semiconductors is not important. For conditions used in our
microscopic calculation, the spin splitting resulting from the Dresselhaus
effect due to the bulk inversion asymmetry (The Dresselhaus splitting is $\sim0.01$~meV in GaAs with doping
density $\sim10^{16}$~cm$^{-3}$), much less than the
detuning of the light from the interband
transitions~\cite{pikus1988} that mediate the effective
interaction, so we can neglect the bulk inversion asymmetry in the
measurement process even though it could be of vital importance in
generating the spin current. Also, in this paper, we consider only
bulk materials, so the structure inversion asymmetry plays no role, though it is the
basis of the Rashba effect. Without considering the
Dresselhaus and Rashba effects due to inversion asymmetries, we avoid
the subtlety in the definition of a spin
current.\cite{sun2005prb,shi2006prl} The effect of inversion asymmetries
on the interaction between the polarized light beams and a spin
current, of course, is worth further study, but we prefer leaving
this question open in this paper.

The paper is organized as follows. Sec.~\ref{Sec_symmetry} presents a systematic
symmetry analysis for the coupling system to give a qualitative
understanding of the linear and circular birefringence effects and the second-order nonlinear optical effect
of pure spin currents. Sec.~\ref{Sec_calculation}
gives the theoretical model and microscopic derivations for both the linear and the second-order nonlinear optical effects, and also explains
the physical pictures for the microscopic mechanism of optical effects of spin currents. Sec.~\ref{Sec_discussion} presents
the numerical results and discussions of the experiment scheme. Sec.~\ref{Sec_conclusion} concludes this paper.

\section{Symmetry Analysis}
\label{Sec_symmetry}
We will particularly consider the
time-reversal ($\mathcal{T}$) and the space-inversion ($\mathcal{P}$) symmetries
of all the relevant physical quantities, and the geometry symmetry for a specific form of spin currents. According to the symmetry
analysis, a pure spin current
may result in a circular birefringence effect but not a linear
birefringence effect, and as it breaks $\mathcal{P}$ symmetry,
a spin current can induce the second-order nonlinear optical processes.

\subsection{Linear optical effects}
We assume the whole system has the $\mathcal{T}$ and $\mathcal{P}$
symmetries at equilibrium. Namely, the effective coupling
between a spin polarization or a spin current
in the  semiconductor system and a probe should have both symmetries, i.e.,
the transformation properties of the effective Hamiltonian $\mathcal {H}_{\mathrm{eff}}$ are
\begin{equation}
\label{eff Ham}
\begin{tabular}{|c|c|c|}
\hline
 & $\mathcal{T}$ & $\mathcal{P}$ \\
\hline $\mathcal {H}_{\text{eff}}$  & + & + \\
\hline
\end{tabular}
\end{equation}
where $+/-$ refers to even/odd under the corresponding symmetry transformations.

In our study, a pure spin current is made of a non-equilibrium
distribution of spin polarization in the momentum space.
In general, it can be quantified by a rank-2 pseudo-tensor defined by (with volume of the material taken as unity)
\begin{equation}\label{spincurrent}
{\mathbb
J}=\sum\limits_{\mathbf{p}}\mathbb{J}_{\mathbf{p}}=e\sum_{\mathbf
p} {\mathbf s}_{\mathbf p}{\mathbf v}_{\mathbf p},
\end{equation}
where ${\mathbf s}_{\mathbf p}$ is the spin polarization and
${\mathbf v}_{\mathbf p}$ is the velocity of a particle with wave vector
${\mathbf p}$, and $e$ is the electron charge. The ``photon spin current'' tensor for a polarized
light beam with electric field
$ {\mathbf F}({\mathbf r},t) =
     \left(F_{+}{\mathbf n}_{+}+F_{-}{\mathbf n}_{-}\right)
     e^{i{\mathbf q}\cdot{\mathbf r}-i\omega_{q}t}    +{\rm c.c.}$
is formulated as
\begin{subequations}
\label{light}
\begin{eqnarray}
{\mathbb I} & \equiv & {\mathbf I}{\mathbf q} \equiv  q\left(I_x{\mathbf x}{\mathbf z}+I_y{\mathbf y}{\mathbf z}
+I_z{\mathbf z}{\mathbf z}\right), \\
I_j & = & \frac{1}{2}
\sum_{\mu,\nu=\pm}\sigma^j_{\mu\nu}F^*_{\mu}F_{\nu},
\end{eqnarray}
\end{subequations}
where ${\mathbf q}$ is the wave vector of the light beam,
the unit vector ${\mathbf z}$ is chosen along the direction of ${\mathbf q}$ so that ${\mathbf q}= q{\mathbf z}$, the unit vectors ${\mathbf x}$ and ${\mathbf y}$ are
related to the light polarization through ${\mathbf n}_{\pm} \equiv
\left(\mp{\mathbf x}-i{\mathbf y}\right)/\sqrt{2}$, and $\sigma^j$ ($j=x, y, z$) is the Pauli matrix.
For completeness, we also consider the spin polarization of the system
\begin{equation}
{\mathbf S}=\sum_{\mathbf p}{\mathbf s}_{\mathbf p}.
\end{equation}

The transformation properties under the $\mathcal{T}$ and $\mathcal{P}$ of the relevant physical quantities are
\begin{equation}\label{TP-sym}
\begin{tabular}{|c|c|c|c|c|c|}
\hline
\ & $\mathbf{S}$ & $\mathbf{q}$  & $\mathbb{J}$ & $I_x,I_y$ &
$I_z$
\\ \hline
$\mathcal{T}$ & $-$ & $-$ & $+$ & $+$ & $-$ \\ \hline
$\mathcal{P}$ & $+$ & $-$ & $-$ & $+$ & $+$ \\ \hline
\end{tabular}
\end{equation}
It is worth mentioning here there is no $|0\rangle$ state in the physical spin of photons, and the photon pseudo spin $I_x$ and $I_y$ do not break the $\mathcal{T}$-symmetry, for it involves the 2nd-order spin flip processes such as $|+1\rangle\rightarrow|0\rangle\rightarrow|-1\rangle$.
In the following we will use these quantities to form an effective Hamiltonian
(undetermined up to a few coupling constants) satisfying the $\mathcal{T}$
and $\mathcal{P}$ symmetries. Since the interaction of the light with
a spin is usually weak, we only consider the effect in the leading order,
which is bilinear in the spin and light quantities.

\emph{Net spin polarization}. The only optical quantity of the same symmetry-breaking type as the spin polarization is $I_z{\mathbf z}$. Thus the
effective interaction between a spin polarization and a light beam has the form
\begin{equation}\label{zero-order}
\mathcal
{H}_{\mathrm{eff}}^{(0)}=\zeta_{0}I_{z}\mathbf{S}\cdot\mathbf{z},
\end{equation}
with a coupling constant $\zeta_0$ to be determined by the specific
microscopic mechanism. Such a coupling corresponds to the
conventional Faraday effect in magnetooptics.\cite{Magnetooptics} We would like to point out here that a spin polarization could induce the Voigt effect. In order to have the same symmetry-breaking type for $I_x$ and $I_y$, the spin polarization should be of an even power. Thus to the leading order, the effective interaction has the form
\begin{align}
\mathcal{H}_{\text{eff}}^{\text{Voigt}}=\zeta_{0}^{\text{Voigt}}I_{x}\left(\mathbf{S}\cdot\mathbf{x}\right)^2.
\end{align}
This explains that the Voigt effect is quadratic in the spin polarization or the applied external magnetic field.

\emph{Pure spin current}. There is no term in the light polarization
$I_j$ ($j=x,y,z$) that has the same symmetry-breaking type as the spin
current, so it is not possible to have linear (nonlinear optics is of course possible) interaction between
the spin current and the light without involving the wave vector.
Considering the wave vector of the light, coupling between the spin
current and the photon current $qI_z{\mathbf {zz}}$  is possible. The linear birefringence effect (similar to the Voigt effect in magnetooptics) is absent. Due to the lack of $|0\rangle$ state in the physical spin of photons, $I_x$ and $I_y$ preserve $\mathcal{T}$ symmetry. Therefore there is no linear coupling of $\mathbb{J}$ to $qI_x\mathbf{xx}$ and $qI_y\mathbf{yy}$.

Furthermore, if the system has spherical symmetry, the effective
Hamiltonian would have a simple tensor contraction form as
\begin{equation}
{\mathcal H}_{\text{eff}}^{(1)}=\zeta_1 q I_z \text{Tr}\left( {\mathbb J} \right)
 +\zeta_2 q I_z {\mathbf z}\cdot {\mathbb J} \cdot {\mathbf z},
\label{Eq_symmetry0}
\end{equation}
with only two coupling constants $\zeta_1$ and $\zeta_2$ to be determined by the microscopic mechanisms.
A possible spherically symmetric system is the vacuum, but in general a semiconductor as a crystal
does not have this symmetry. The general effective interaction in a semiconductor should have the form
\begin{equation}
{\mathcal H}_{\text{eff}}^{(1)}=qI_z{\mathbf z}{\mathbf z}:\mathcal{A}:{\mathbb J},
\end{equation}
where $\mathcal{A}$ is a parameter tensor determined by the microscopic structure of the material.
Since only the light polarization term $I_z$ appears in the interaction,
the optical birefringence effect is circular, similar to the Faraday rotation.

In realistic case, the spin current often has some special form. As a general case
a spin current tensor can have the form as
\begin{equation}
\mathbb{J}=J_X\mathbf{XZ}+J_Y\mathbf{YZ}+J_Z\mathbf{ZZ}=\mathbf{JZ},
\end{equation}
where ${\mathbf Z}$ is the unit vector along the direction of spin current, the unit vectors ${\mathbf X}$ and ${\mathbf Y}$ are perpendicular to ${\mathbf Z}$, and ${\mathbf J}$ denotes the spin current amplitude vector, which is an axial vector parallel to the spin polarization direction.
Now the ${\mathbf z}$ and ${\mathbf Z}$ axes form a special plane. If the system has reflection symmetry
with respect to this plane (e.g., the system is spherically symmetric or the plane is along a special
crystal direction of the semiconductor),  the symmetry properties of
the relevant quantities under reflection with respect to the ${\mathbf z}$-${\mathbf Z}$ plane
will impose further constraint on the interaction and significantly simplify the Hamiltonian.
Under the reflection, the relevant quantities transform as
\begin{equation}
\begin{tabular}{|c|c|c|c|c|}
\hline
 & \ \ q  & $I_z$ & $\mathbf{J}_{\parallel}$ & $\mathbf{J}_{\perp}$ \\
\hline
Reflection with ${\mathbf z}$-${\mathbf Z}$ plane & $+$  & $-$ & $-$ & $+$ \\
\hline
\end{tabular}
\end{equation}
where ${\mathbf J}_{\parallel}$ is the component of ${\mathbf J}$ in the plane and
${\mathbf J}_{\perp}$ is the perpendicular component. By the table above, it is evident that to keep the effective Hamiltonian invariant only the in-plane component of ${\mathbf J}_{\parallel}$ would couple with the $q I_z$. Without loss of generality, let ${\mathbf Y}$ be perpendicular to the
${\mathbf z}$-${\mathbf Z}$ plane, the effective Hamiltonian reads
\begin{equation}
{\mathcal H}_{\text{eff}}^{(1)} = A_1 q I_z J_Z+A_2 qI_zJ_X,
\end{equation}
in which two coupling constants $A_1$ and $A_2$ are to be determined by microscopic calculation.
Alternatively, the Hamiltonian can be expressed in a form independent of the choice of the ${\mathbf X}$ and ${\mathbf Y}$ axes as
\begin{equation}
{\mathcal H}_{\text{eff}}^{(1)} = \zeta_1 qI_zJ_Z+\zeta_2 q I_z\mathbf{z}\cdot\mathbb{J}\cdot\mathbf{z},
\end{equation}
which is the same as Eq.~(\ref{Eq_symmetry0}), but does not require the spherical symmetry of materials.

The physical effect of the effective coupling can be extracted from
the linear optical susceptibility,
\begin{align}
\chi_{\mu,\nu}+\chi^*_{\nu,\mu} &= (1/\epsilon_0){\partial^2
{\mathcal H}_{\rm eff}}/({\partial F^*_{\mu}\partial F_{\nu}}),
\label{chi}
\end{align}
where $\epsilon_0$ is the vacuum permittivity. Thus we get an
opposite susceptibility for opposite circular polarization in
presence of a spin polarization or a pure spin current
\begin{subequations}
\begin{align}
\chi^{(0)}_{++} &= -\chi^{(0)}_{--}=(1/4\epsilon_0)\zeta_0\mathbf{z\cdot
S},\label{FR0}
\\
\chi^{(1)}_{++} &= -\chi^{(1)}_{--}=(q/4\epsilon_0)\left(\zeta_1
J_Z + \zeta_2{\mathbf z}\cdot\mathbb{J}\cdot{\mathbf z}
       \right).
\label{FR1}
\end{align}
\end{subequations}
The effective energy shift resulting from the light-spin or light-spin current interaction means a phase shift in the light observed in the far-field. Eq.~(\ref{FR0}) is nothing but the conventional Faraday rotation in
magnetooptics,\cite{Magnetooptics} Eq.~(\ref{FR1}) indicates that a pure spin current would produce a circular birefringence effect. This new effect of a pure spin current may be dubbed ``\emph{spin current Faraday effect}''\cite{wang2008prl} because of its similarity to the conventional Faraday rotation due to magnetization, with awareness that a pure spin current, however, bears no net magnetization.

\subsection{Second-order nonlinear optical effects}
\label{sec_second}
The second-order nonlinear optical effect such as sum-frequency process is characterized by a second-order nonlinear susceptibility $\chi^{(2)}$ via
\begin{align}\label{SFG_definition}
\mathbf{P}^{(2)}(\omega_1+\omega_2) &= \chi^{(2)}:\mathbf{F}_1(\omega_1)\mathbf{F}_2(\omega_2),
\end{align}
where $\mathbf{F}_1$ and $\mathbf{F}_2$ are the two optical fields, $\mathbf{P}$ is the induced polarization, and $\chi^{(2)}$ is a rank-3 tensor. Under $\mathcal{P}$ operation, $\mathbf{F}_1$, $\mathbf{F}_2$ and $\mathbf{P}$ reverse the sign, which means $\chi^{(2)}$ is zero if the system has $\mathcal{P}$-symmetry. A pure spin current breaks the $\mathcal{P}$-symmetry, results in a nonzero $\chi^{(2)}$, and makes the second-order nonlinear optical process possible.

In general, as a rank-3 tensor $\chi^{(2)}$ has 27 independent components
\begin{align}
\chi^{(2)} &= \chi_{XXX}\mathbf{XXX}+\chi_{XXY}\mathbf{XXY}+\ldots+\chi_{ZZZ}\mathbf{ZZZ}.
\end{align}
But the symmetry properties of the spin current and the system will impose constraints on $\chi^{(2)}$, reducing the number of independent parameters.\cite{YRShen} For a longitudinal spin current $J_Z\mathbf{ZZ}$, in which the spin polarization is parallel or antiparallel to the current direction, the spin current is reversed under the reflection with respect to the $X$-$Z$ plane, so that $\chi_{XXX}\mathbf{XXX}+\chi_{XXY}\mathbf{XX}(-\mathbf{Y})+\ldots+\chi_{ZZZ}\mathbf{ZZZ}=-\chi^{(2)}$ and the terms with direction $\mathbf{Y}$ appeared even times (twice or zero times) must vanish. Similarly, the longitudinal spin current is reversed under the reflection with respect to the $Y$-$Z$ and $X$-$Y$ planes, so the terms with direction $\mathbf{X}$ or $\mathbf{Z}$ that appear even times must be zero. Moreover, the longitudinal spin current should be invariant under the $\pi/2$ rotation with respect to its current direction, therefore $\chi_{XYZ}=-\chi_{YXZ}$, $\chi_{YZX}=-\chi_{XZY}$, $\chi_{ZXY}=-\chi_{ZYX}$. With all these constraints, the sum-frequency susceptibility induced by a longitudinal spin current $J_Z\mathbf{ZZ}$ can be expressed as
\begin{align}\label{chi2Jz}
\chi^{(2)}_{J_Z} = & J_Z\left[\alpha_1(\mathbf{XYZ}-\mathbf{YXZ}) + \alpha_2(\mathbf{YZX}-\mathbf{XZY})\nonumber\right.
\\
& \left. + \alpha_3(\mathbf{ZXY}-\mathbf{ZYX})\right],
\end{align}
where there are only three independent parameters $\alpha_i$ ($i=1,2,3$). For a transverse spin current $J_X\mathbf{XZ}$, in which the spin polarization is perpendicular to the current direction, the current is reversed under reflection with respect to the $X$-$Z$ plane, but invariant under refection with respect to $X$-$Y$ or $Y$-$Z$ plane. Then the terms that contain even times of $\mathbf{Y}$ or odd times of $\mathbf{Z}$ or $\mathbf{X}$ must be zero, so
\begin{align}\label{chi2Jx}
\chi^{(2)}_{J_X} = & J_X\left(x_1\mathbf{XXY}+x_2\mathbf{XYX}+x_3\mathbf{YXX}+z_1\mathbf{ZZY}\right.\nonumber
\\
&\left.+z_2\mathbf{ZYZ}+z_3\mathbf{YZZ}+y\mathbf{YYY}\right),
\end{align}
with seven independent parameters to be determined. Similar symmetry analysis can be applied to $J_Y\mathbf{YZ}$. Such unique polarization dependence of the second-order optical susceptibility can be used to distinguish the longitudinal and transverse components of a spin current, and also to single out the spin-current signature from the effects of the material background or a charge current.\cite{khurgin1995apl,ruzicka2012prl}

\section{Microscopic calculation}
\label{Sec_calculation}
To quantitatively determine the linear and the second-order nonlinear optical effects of a spin current, we will perform the microscopic calculation for a pure spin current in a bulk direct-gap semiconductor using the standard perturbation theory.\cite{YRShen,sipe2000prb} We employ the eight-band model.\cite{Yu} We assume that the pure spin current result from a non-equilibrium distribution of electrons in the conduction band. Namely,  as shown in Fig.~\ref{fig1}(a), a small portion of non-equilibrium electrons with opposite velocities near the Fermi surface have opposite spin polarizations, which is similar to the situation in Ref.~\onlinecite{kato2004science}. The optical interaction includes the interband transitions and the intraband acceleration of electrons and holes. To avoid real absorption of light, the light frequencies are chosen to be below the absorption edge in linear optical effect, and the sum frequency is below the two-photon absorption edge in the second-order nonlinear optical effect.

\subsection{Model}
\label{Sec_Model}
\begin{figure}[t]
\includegraphics[width=3.4in,clip=true]{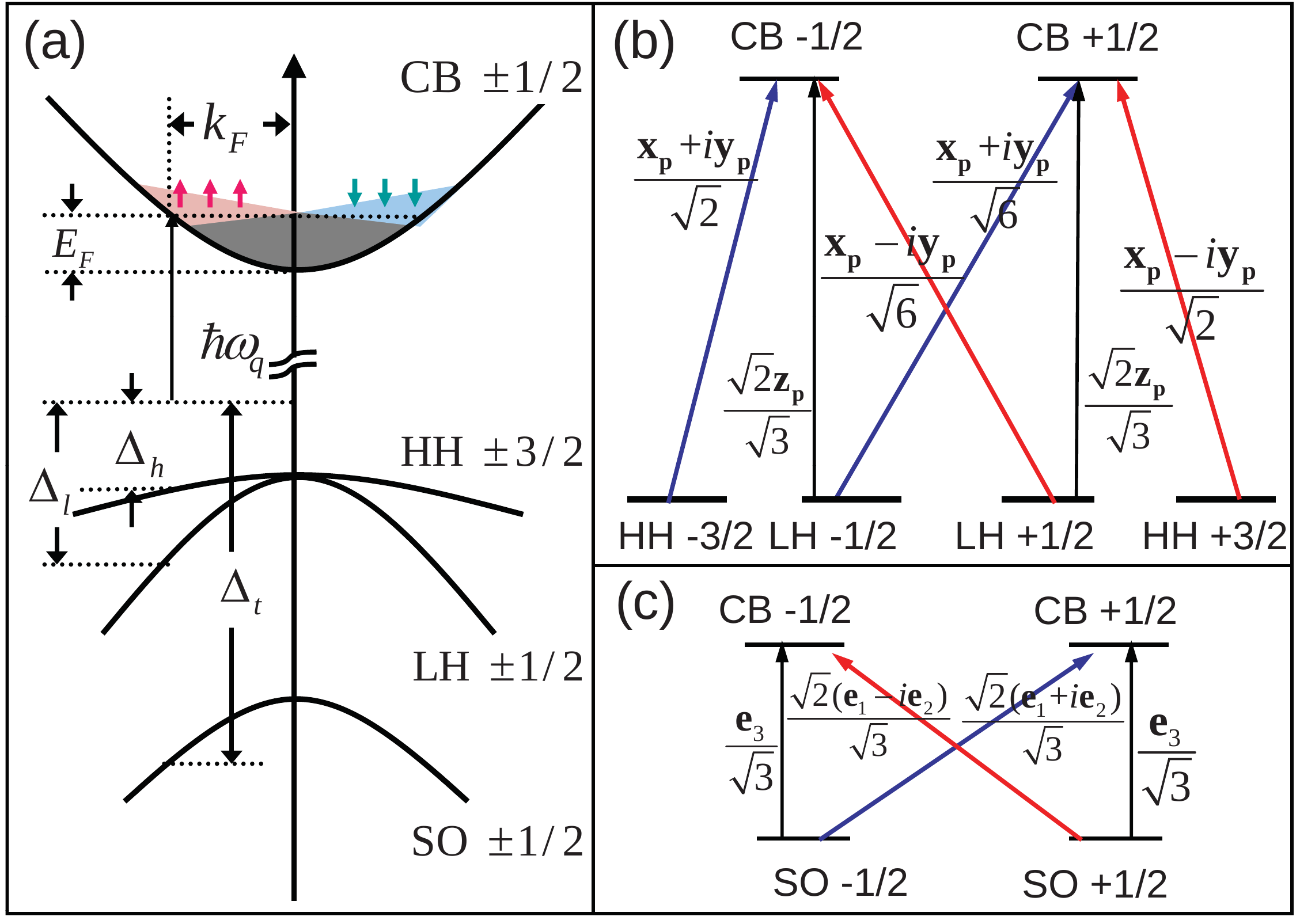}
\caption{(color online) (a) Schematic band structure of the eight-band model near the
$\Gamma$ point of an n-doped III-V compound semiconductor, and
illustration of the spin non-equilibrium distribution of electrons for a pure spin current. (b) \& (c) Selection rules and relative dipole moments from the HH and LH, SO bands to the CB.}
\label{fig1}
\end{figure}

We consider an n-doped direct-gap semiconductor of GaAs as a model material. Since other bands are separated far away in energy, we assume the near-gap optical interactions in GaAs involve mostly the eight bands around the fundamental gap, including the conduction band (CB), the heavy-hole (HH) band, the light-hole (LH) band, and the spin-orbit split-off (SO) band, each of 2-fold degeneracy [Fig.~\ref{fig1}(a)].

Near the $\Gamma$-point of the Brillouin zone, the energy dispersion of the CB and the SO electron is almost parabolic and isotropic, which  can be respectively written as $E_{e{\mathbf
p}}=p^2/(2m_e)+E_g$ and $E_{t{\mathbf
p}}=p^2/(2m_{t})+E_{\text{SO}}$, where $m_{e/t}$ is the effective mass of the CB/SO band, $E_g$ is the fundamental band gap with the energy zero at the top of the valence band, and $E_{\text{SO}}$ is the split-off energy due to the spin-orbit coupling. Hereafter the Planck constant $\hbar$ is taken as unity. For the HH and LH bands, since the energy band anisotropy will not affect the symmetry analysis as shown in Sec.~\ref{Sec_symmetry} ( The 4-fold rotation symmetry will be retained even considering the anisotropic dispersion), we will neglect the anisotropy effect in this Section, and take it into account separately in Sec.~\ref{anisotropy}. Thus we express the \emph{isotropic} Luttinger-Kohn Hamiltonian $H^I_{\rm LK}$ for the HH and LH bands near the band edge as\cite{HandbookOnSemiconductors}
\begin{align}\label{H_L}
H^{I}_{\rm LK} = \frac{1}{2m_0}\left[\left(\gamma_1
+\frac{5}{2}\gamma_2\right){\boldsymbol \nabla}^2
  - 2\gamma_2\left({\boldsymbol \nabla} \cdot {\mathbf K}\right)^2\right],
\end{align}
where ${\mathbf K}$ is a spin-3/2 for the total angular momentum of
an electron in the HH and LH bands, $\gamma_1$ and $\gamma_2$ are the Luttinger
parameters, and $m_0$ is the free electron mass.  The isotropic Luttinger-Kohn Hamiltonian
can be diagonalized with the spin-$3/2$ quantized along the direction of the momentum $\mathbf{p}$. The HH band with magnetic quantum numbers $\pm3/2$ has the energy dispersion $E_{h\mathbf{p}}=(\gamma_1-2\gamma_2)p^2/2m_0\equiv p^2/2m_h$; and the LH band has magnetic quantum numbers $\pm1/2$ with the dispersion relation as $E_{l\mathbf{p}}=(\gamma_1+2\gamma_2)p^2/2m_0\equiv p^2/2m_l$, where $m_{h/l}$ is the effective mass of the HH/LH band. If the HH-LH splitting is neglected further, the HH and LH bands become a $4$-fold degenerate spin-$3/2$ band and the spin quantization direction can be chosen independent of the momentum.

The Bloch state of a CB electron with momentum $\mathbf{p}$ is $|\psi^c_{\pm}(\mathbf{p})\rangle\equiv\hat{e}^{\dag}_{\pm,\mathbf{p}}|0\rangle=e^{i\mathbf{p\cdot r}}|\pm\rangle_{\bf p}$, where $|0\rangle$ represents the vacuum state and $\hat{e}^{\dag}_{\pm,\mathbf{p}}$ denotes an creation operator that produces an electron in CB with spin
$\pm 1/2$ and momentum $\mathbf{p}$. The Bloch state of an electron in the valence band with momentum $\bf p$ is  $|\psi^{\alpha}_{m}(\mathbf{p})\rangle\equiv\hat{V}^{\dag}_{j,m;\mathbf{p}}|0\rangle=e^{i\mathbf{p\cdot r}}|j,m\rangle_{\bf p}$, in which
$j=3/2$ and $m=\pm3/2$ stands for the HH band ($\alpha=h$), $j=3/2$ and $m=\pm1/2$ for the LH band ($\alpha=l$), $j=1/2$ and $m=\pm1/2$ for the SO band ($\alpha=t$), and $\hat{V}_{j,m;\mathbf{p}}$ denotes the annihilation operator for an electron in the corresponding valence band.
Then the non-interacting Hamiltonian is
\begin{align}
\hat{H}_0 = & \sum_{\mu=\pm,{\mathbf p}}
\left( E_{e{\mathbf p}}\hat{e}_{\mu,{\mathbf p}}^{\dagger}\hat{e}_{\mu,{\mathbf p}}
+ E_{h{\mathbf p}}\hat{h}_{\mu,{\mathbf
p}}^{\dagger}\hat{h}_{\mu,{\mathbf p}} + E_{l{\mathbf
p}}\hat{l}_{\mu,{\mathbf p}}^{\dagger}\hat{l}_{\mu,{\mathbf p}}\right.
\nonumber
\\
&\left. +
E_{t{\mathbf p}}\hat{t}_{\mu,{\mathbf
p}}^{\dagger}\hat{t}_{\mu,{\mathbf p}} \right),
\end{align}
where the hole operators are defined as $\hat{h}_{\mp,-{\mathbf p}} \equiv \hat{V}^{\dag}_{3/2,\pm
3/2; {\mathbf p}}$, $\hat{l}_{\mp,-{\mathbf p}} \equiv \hat{V}^{\dag}_{3/2,\pm
1/2; {\mathbf p}}$, and $\hat{t}_{\mp,-{\mathbf p}} \equiv \hat{V}^{\dag}_{1/2,\pm
1/2; {\mathbf p}}$.
It should be pointed out that here
the angular momentum ${\mathbf K}$ is quantized along ${\mathbf p}$
so that the spin-orbit coupling in the valence bands has already been included.

The initial state of the system (before optical excitation) is characterized by a density matrix $\hat{\rho}_0$. We assume that the system has translation symmetry and initially there is no hole in the system, so we have
\begin{subequations}
\begin{align}
&{\rm Tr}\left[\hat{\rho}_0\hat{h}^{\dag}_{\mu{\mathbf
k}}\hat{h}_{\nu{\mathbf k'}}\right] ={\rm
Tr}\left[\hat{\rho}_0\hat{l}^{\dag}_{\mu{\mathbf
k}}\hat{l}_{\nu{\mathbf k'}}\right]={\rm
Tr}\left[\hat{\rho}_0\hat{t}^{\dag}_{\mu{\mathbf
k}}\hat{t}_{\nu{\mathbf k'}}\right]=0,
\\
&{\rm Tr}\left[\hat{\rho}_0\hat{e}^{\dag}_{\mu{\mathbf
k}}\hat{e}_{\nu{\mathbf k'}}\right]
=\delta_{\mathbf{k,k'}}f_{\mu\nu,{\mathbf k}}
.
\end{align}
\end{subequations}
The spin current, which results from the non-equilibrium distribution of CB electrons, is expressed by Eq.~(\ref{spincurrent}), where the velocity and the spin polarization of an electron with momentum
$\mathbf{p}$ is respectively given by $\mathbf{v_p}=\nabla_{\mathbf{p}}E_{e\mathbf{p}}$ and  $\mathbf{s_p}=(1/2)\sum_{\mu\nu}\boldsymbol{\sigma}_{\mu\nu}f_{\mu\nu,\mathbf{p}}$ with
$\boldsymbol{\sigma}$ denoting the Pauli matrices.

\subsection{Linear optical effects}

The direct interaction between a light beam and a semiconductor is the dipole \emph{interband} optical transitions. Only through the spin-orbit coupling in valence bands, may the light beam interact with the spin of electrons.

For the dipole interband transition [Fig.~\ref{fig1}(b)\&(c)], the polarization density operator reads\cite{sipe2000prb}
\begin{align}\label{interband}
\hat{\mathbf{P}}_{\text{inter}}(\mathbf{r})  = & - d_{\text{cv}}^*\sum\limits_{\mathbf{k,p};\mu=\pm}\left(
 \mathbf{n}_{\bar{\mu},{{\mathbf p}}}\hat{h}_{\bar{\mu},-{\mathbf p}}\hat{e}_{\mu,{\mathbf k}}
+{\frac{1}{\sqrt{3}}}\mathbf{n}_{{\mu},{{\mathbf
p}}}\hat{l}_{\mu,-{\mathbf p}}\hat{e}_{{\mu},{\mathbf k}}\right.
\nonumber
\\
&\left.-\sqrt{\frac{2}{3}}{\mathbf z}_{{\mathbf
p}}\hat{l}_{\bar{\mu},-{\mathbf p}}\hat{e}_{\mu,{\mathbf k}}
-\mu\sqrt{\frac{2}{3}}\mathbf{n}_{{\mu},{{\mathbf
p}}}\hat{t}_{\mu,-{\mathbf p}}\hat{e}_{{\mu},{\mathbf k}}\right.
\nonumber
\\
&\left.
+{\frac{\mu}{\sqrt{3}}}{\mathbf z}_{{\mathbf
p}}\hat{t}_{\bar{\mu},-{\mathbf p}}\hat{e}_{\mu,{\mathbf k}} \right)
e^{i\mathbf{p\cdot r}-i\mathbf{k\cdot r}} +
\mathrm{H.c.},
\end{align}
where ${\mathbf
n}_{\pm,{\mathbf p}}\equiv\mp\left({\mathbf x}_{\mathbf p}\pm
i{\mathbf y}_{\mathbf p}\right)/\sqrt{2}$ denotes the right/left
circular polarization about the momentum direction ${\mathbf p}$, ${\mathbf z}_{\mathbf p}\equiv {\mathbf p}/p$, and $\bar{\mu}\equiv -{\mu}$. As will be discussed in Sec,~\ref{anisotropy}, the momentum dependence of the dipole moment has no significant effect, so here we assume the interband dipole moment $d_{\text{cv}}$ independent of the momentum. With the dipole interaction with a light $\hat{H}_1(t) = -\int \hat{{\mathbf P}}_{\text{inter}}({\mathbf r})\cdot{\mathbf F}({\mathbf r},t)d\mathbf{r}$,
the light-matter interaction Hamiltonian in the rotating wave approximation can be explicitly expressed as
\begin{align}
\hat{H}_1
\equiv&\exp\left(\sum\limits_{\mu,\mathbf{k}}i\omega_{q}t\hat{e}^{\dag}_{\mu,\mathbf{k}}\hat{e}_{\mu,\mathbf{k}}\right)\hat{H}_1(t)
\exp\left(\sum\limits_{\mu,\mathbf{k}}-i\omega_{q}t\hat{e}^{\dag}_{\mu,\mathbf{k}}\hat{e}_{\mu,\mathbf{k}}\right)
\nonumber
\\
=& {d_{\rm cv}^*}\sum_{\mu,\nu,{\mathbf p}}
  F^*_{\nu}{\mathbf n}_{\nu}^*\cdot
  \left({\mathbf n}_{\bar{\mu},{\mathbf p}}\hat{h}_{\bar{\mu},-{\mathbf p}}\hat{e}_{\mu,{\mathbf q}+{\mathbf p}}
 + \frac{1}{\sqrt{3}}{\mathbf n}_{\mu,{\mathbf p}}\hat{l}_{\mu,-{\mathbf p}}\hat{e}_{\mu,{\mathbf q}+{\mathbf p}}\right.
 \nonumber
 \\
 &\left. -\sqrt{\frac{2}{3}}{\mathbf z}_{\mathbf p}\hat{l}_{\bar{\mu},-{\mathbf p}}\hat{e}_{\mu,{\mathbf q}+{\mathbf p}}
 -\mu\sqrt{\frac{2}{3}}{\mathbf n}_{\mu,{\mathbf p}}\hat{t}_{\mu,-{\mathbf p}}\hat{e}_{\mu,{\mathbf q}+{\mathbf p}}\right.
  \nonumber \\
&\left. +\frac{\mu}{\sqrt{3}}{\mathbf z}_{{\mathbf p}}\hat{t}_{\bar{\mu},-{\mathbf p}}\hat{e}_{\mu,{\mathbf q}+{\mathbf p}}\right)
  +{\rm H.c.}.
\end{align}

Under the condition that the optical interaction strength is much
smaller than the detuning of the light from the valence band to the Fermi level (the perturbation regime),
the effective energy due to the dipole interaction can be derived by the second-order perturbation as
\begin{equation}\label{eff_formula}
{\mathcal H}_{\text{eff}} =  {\rm Tr}\left[\hat{\rho}_0 \hat{H}_1
\left(\hat{H}_0-\omega_q\right)^{-1} \hat{H}_1\right].
\end{equation}
Such effective coupling between a spin current and a polarized light beam on the one hand can be regarded as the
frequency shift of the light in the presence of the spin current, and on the other hand can be
considered as the energy change of the semiconductor system under the driving of light beam.
The second-order perturbation means that there are two virtual optical transitions induced by the electric field of the light: one creating an electron-hole pair and one annihilating the electron-hole pair. The virtual excitations cause no real optical
absorption but a phase shift, indicating that the effective coupling is real.
The optical effect of the spin-current can be understood as the Pauli
blocking in the transition involving different spin states. With this picture in mind,
the following microscopic calculation, though lengthy, is quite transparent.

\subsubsection{Physical Picture}
The physical picture for the microscopic mechanism of the spin current Faraday effect is rooted in the fact that a spin will induce a Faraday rotation like a magnet. In Faraday rotation, a linearly polarized optical field ${\mathbf F}$ induces a polarization as a rotation about the spin,
\begin{align}
\mathbf{P}^{(1)} &\propto \frac{\mathbf{F}\times\mathbf{s_k}}{\omega-E_{\mathbf{k}}},
\end{align}
where $\mathbf{s_k}$ is the spin polarization associated with the state of $\mathbf{k}$, $E_{\mathbf{k}}$ is the resonant optical transition energy. This naturally explains Faraday rotation due to spin polarization as in Eq.~(\ref{effective_0}).

\begin{figure}[t]
\includegraphics[width=2.9in,clip=true]{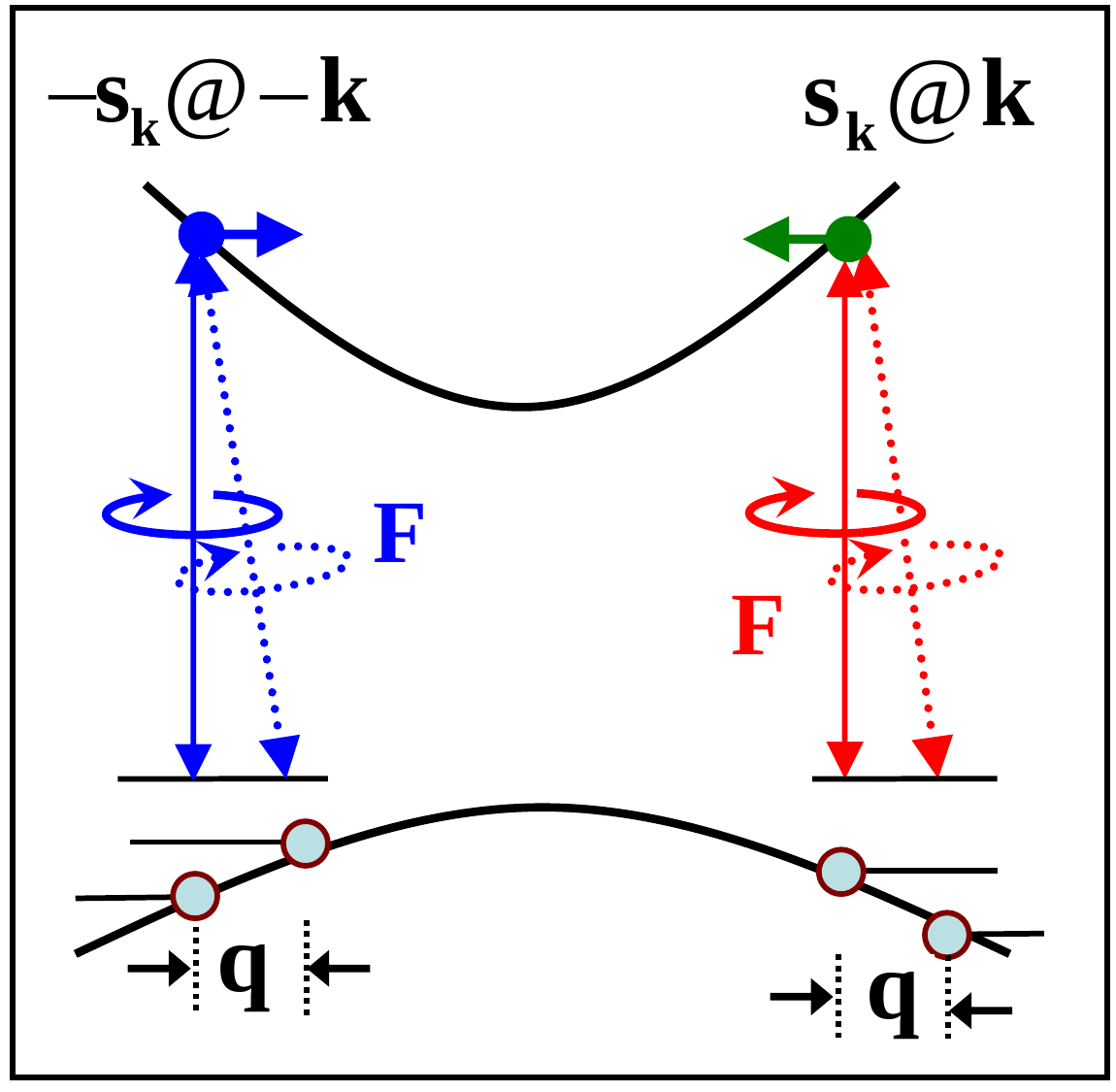}
\caption{(color online) Physical picture for the microscopic mechanism of the \emph{spin current Faraday effect}. The virtual transition energy is the same for $\pm\mathbf{s_k}$ when neglecting light momentum $\mathbf{q}$.}
\label{fig2}
\end{figure}

For a pure spin current, we first consider only a pair of spins, $\mathbf{s_k}$ at momentum $\mathbf{k}$, and $-\mathbf{s_k}$ at momentum $-\mathbf{k}$ in the CB [Fig.~\ref{fig2}]. This pair can be viewed as a generator of pure spin current. $\mathbf{s_k}$ gives rise to a Faraday rotation of $\mathbf{P}^{(1)}_{\mathbf{k}}\propto\mathbf{F}\times\mathbf{s_k}/(\omega-E_{+\mathbf{k}})$; while $-\mathbf{s_k}$ leads to a Faraday rotation of $\mathbf{P}^{(1)}_{-\mathbf{k}}\propto\mathbf{F}\times\mathbf{s_k}/(\omega-E_{-\mathbf{k}})$. Therefore, the Faraday rotations caused by the pair of spins cancel each other in the vertical optical transition. However, when the effect of the small light-momentum is taken into consideration, the excitation energy at $\pm\mathbf{k}$ will shift respectively to $E_{\pm\mathbf{k}}\rightarrow E_{\mathbf{q\pm k}}\approx E_{\pm\mathbf{k}}\pm\mathbf{q\cdot v_k}$ [Fig.~\ref{fig2}], and $\pm\mathbf{s_k}$ will induce different Faraday rotations due to opposite energy variation. Up to the first order of $\mathbf{q}$, the polarization is $\mathbf{P}^{(1)}\propto\mathbf{F}\times\mathbf{s_kv_k}\cdot\mathbf{q}/(\omega-E_{\mathbf{k}})^2$, where $e\mathbf{s_kv_k}$ is just the spin current tensor contributed by the pair of electrons. This explains the $q$-dependence of the spin current Faraday effect. More generally, the hole state wavefunction is also changed when considering the light momentum, which causes extra Berry phase effects [terms proportional to $1/E_F$ in Eq.~(\ref{couplingconstatnt})].

\subsubsection{Effective Hamiltonian by SO-CB transitions}
\label{Effctive Hamiltonian}
To better understand the microscopic mechanism of the light-spin current coupling, let us first derive the effective Hamiltonian contributed solely by the transitions between the CB and SO bands. The SO band electrons has 2-fold degeneracy, and the spin states as well as the selection rules for the interband transitions, like the CB electrons, are independent of the momentum [Fig.~\ref{fig1}(c)].

We first consider a single electron with momentum ${\bf k}$ and spin polarization
${\bf s}_{\bf k}$. The spin current contributed by this electron
is ${\mathbb J}_{\bf k}=e{\bf s}_{\bf k}{\bf v}_{\bf k}$
with the velocity ${\bf v}_{\bf k}=\bf{k}/ m_e$.
It is convenient to define the spin basis states along the spin polarization direction.
In such chosen basis, the spin density matrix of the electron is diagonal. With the population
in the spin-up and spin-down states denoted as $f_{+}$ and $f_{-}$, respectively, the spin polarization
is $s_{\bf k}=(f_+-f_-)/2$. The interband transitions $|1/2,\pm 1/2\rangle_{\mathbf k}\leftrightarrow |\mp\rangle_{\mathbf k}$
couple to a field with circular polarization $\mp({\bf e}_1\pm i{\bf e}_2)/\sqrt{3}$,
and the vertical inter-band transitions $|1/2,\pm 1/2\rangle_{\bf k}\leftrightarrow |\pm\rangle_{\bf k}$
couple to a field of linear polarization ${\bf e}_3/\sqrt{3}$, where the coordinate system is so defined that
${\bf e}_3$ is along the spin polarization direction of the electron considered. Summing up all possible
inter-band transitions, the energy shift of this electron due to coupling to an optical field
${\bf F}$ is
\begin{align}
{\mathcal H}^{\rm SO}_{\rm eff, {\mathbf k}} = &
-\frac{1}{3}\left|d_{\mathrm{cv}}\right|^2
\sum\limits_{\pm}
\frac{
\left(1-f_{\pm}\right){\mathbf F}^*\cdot \left(\mathbf{e}_1\mp
i\mathbf{e}_2\right)\left(\mathbf{e}_1\mp
i\mathbf{e}_2\right)^*\cdot\mathbf{F}
}
{\omega_q-E_{t,-\mathbf{k}+{\mathbf q}}-E_{e\mathbf k}}
\nonumber
\\
&-\frac{1}{3}\left|d_{\mathrm{cv}}\right|^2
\sum\limits_{\pm}
\frac{
\left(1-f_{\pm}\right){\mathbf F}^*\cdot \mathbf{e}_3\mathbf{e}_3^*\cdot\mathbf{F}
}
{\omega_q-E_{t,-\mathbf{k}+{\mathbf q}}-E_{e\mathbf k}}
,
\end{align}
where the factor $(1-f_{\pm})$ accounts for the Pauli blocking of the interband transitions.
The second term, which is related to vertical transitions caused by a
linearly polarized field, does not depend on the spin polarization, so it can be dropped
as the background. With expansion to the first order of ${\mathbf q}$ and omission of the background terms, the energy shift becomes
\begin{align}
{\mathcal H}^{\rm SO}_{\rm eff, {\mathbf k}} = & +\frac{2}{3}i\left|d_{\text{cv}}\right|^2
\frac{
s_{\mathbf k}{\mathbf F}^*\cdot \left(\mathbf{e}_1\mathbf{e}_2-\mathbf{e}_2
\mathbf{e}_1\right)\cdot\mathbf{F}}
{\omega_q-E_{t,-\mathbf{k}}-E_{e\mathbf k}}\nonumber
\\
& +\frac{2}{3}i\left|d_{\mathrm{cv}}\right|^2
\frac{
s_{\mathbf k}{\mathbf F}^*\cdot \left(\mathbf{e}_1\mathbf{e}_2-\mathbf{e}_2
\mathbf{e}_1\right)\cdot\mathbf{F}
{\mathbf q}\cdot{\mathbf k}}
{m_t(\omega_q-E_{t,-\mathbf{k}}-E_{e\mathbf k})^2}.
\end{align}
Since $\left(\mathbf{e}_1\mathbf{e}_2-\mathbf{e}_2
\mathbf{e}_1\right)\cdot\mathbf{F}=\mathbf{F}\times\left(\mathbf{e}_1\times\mathbf{e}_2\right)=\mathbf{F}\times\mathbf{e}_3$, the physical meaning of this coupling is transparent: the linear-polarized optical field will tilt about
the spin, which is essentially the Faraday rotation with spin playing the role of a magnet. The summation over the momentum space gives
\begin{align}\label{SO}
{\mathcal H}^{\rm SO}_{\rm eff} &=
-\frac{4}{3}\left|d_{\rm cv}\right|^2\frac{1}{\Delta_t}I_z\mathbf{z\cdot S}
-\frac{4}{3}\left|d_{\rm cv}\right|^2\frac{m_e}{em_t\Delta_t^2}qI_z\mathbf{z}\cdot\mathbb{J}\cdot\mathbf{z},
\end{align}
where $\Delta_t$ is the light detuning from SO band to the Fermi surface.

\subsubsection{Effective coupling by transitions between HH/LH and CB}
If the HH bands and LH bands are assumed degenerate, the quantization direction of the $3/2$-spin of the HH and LH can be chosen arbitrarily and the effective Hamiltonian are obtained in a similar way to that contributed by the SO-CB transition. However, with the HH-LH splitting considered, the quantization direction of the hole states depends on its momentum, thus, with the trivial background omitted, the effective Hamiltonian can be derived explicitly as
\begin{subequations}
\begin{align}
{\mathcal H}^{\rm HL}_{\rm eff} = & {|d_{\mathrm{cv}}|^2} \left[
I_x\left(\mathbf{xx-yy}\right)+I_y\left(\mathbf{xy+yx}\right)+I_z\left(i\mathbf{xy}-i\mathbf{yx}\right)\right]:
\nonumber\\
& \sum_{\mu=\pm,{\mathbf p}}\Bigg[f_{\bar{\mu}_{\mathbf
p}\bar{\mu}_{\mathbf p}, {\mathbf q}+{\mathbf p}}
        \frac{{\mathbf n}_{\mu,{\mathbf p}}{\mathbf n}^*_{\mu,{\mathbf p}}}
         {E_{e{\mathbf q}+{\mathbf p}}+E_{h-{\mathbf p}}-\omega_q}\label{HL1}
         \\
& +\frac{1}{3}f_{{\bar{\mu}}_{\mathbf p}{\bar{\mu}}_{\mathbf p},
{\mathbf q}+{\mathbf p}}
        \frac{{\mathbf n}_{\bar{\mu},{\mathbf p}}{\mathbf n}^*_{\bar{\mu},{\mathbf p}}+2{\mathbf z}_{\mathbf p}{\mathbf z}_{\mathbf p}^*}
         {E_{e{\mathbf q}+{\mathbf p}}+E_{l-{\mathbf p}}-\omega_q}\label{HL2}
         \\
& -\frac{\sqrt{2}}{3}f_{\mu_{\mathbf p}\bar{\mu}_{\mathbf
p}, {\mathbf q}+{\mathbf p}}
                    \frac{{\mathbf n}_{\bar{\mu},{\mathbf p}}{\mathbf z}_{\mathbf p}^*+{\mathbf z}_{\mathbf p}{\mathbf n}_{{\mu},{\mathbf p}}^*}
                     {E_{e{\mathbf q}+{\mathbf p}}+E_{l-{\mathbf
                     p}}-\omega_q}\Bigg],\label{HL3}
\end{align}
\end{subequations}
where $\mu_{\mathbf p}$ indicates the spin moment quantized along
${\mathbf p}$. The terms in the equation above stand for different physical processes as follow.
Term (\ref{HL1}) accounts for the HH-CB transitions where an (virtually)
absorbed photon will be emitted with the same circular
polarization conserving the electron spin. The
LH-CB optical transition can be either circularly polarized or linearly polarized. In
(\ref{HL2}), the absorbed and emitted photons in the virtual LH-CB transitions have the same polarization with the electron spin
conserved. In (\ref{HL3}), the optical polarizations involved in the LH-CB absorption and emission are changed, leading to an angular momentum transfer between the light and the electron, while the total angular momentum
is still conserved.

The spin-independent population terms such as $f_{++,{\mathbf
p}}+f_{--,{\mathbf p}}$ are related to a change in the background refraction index, but not to the spin polarization and spin current, we will drop the spin-independent population terms and keep the spin polarization terms $\mathbf{s_p}$ only in the effective coupling as
\begin{widetext}\begin{align}\label{anti-tensor}
{\mathcal H}^{\text{HL}}_{\rm eff} = & \left|d_{\mathrm{cv}}\right|^2 \Big[
I_x\left(\mathbf{xx-yy}\right)+I_y\left(\mathbf{xy+yx}\right)+iI_z\left(\mathbf{xy}-\mathbf{yx}\right)
\Big]
:i\sum\limits_{\mathbf{p}}\left[
\left(\frac{f_{\mathbf{z_{p}},\mathbf{p+q}}(\mathbf{x_py_p-y_px_p})}{E_{e,{{\mathbf
q}+{\mathbf p}}}+E_{h,\mathbf{p}}-\hbar\omega_q} - \frac{1}{3}\left(E_h\rightarrow E_l\right)\right)
+\frac{2}{3}\times\right.
\nonumber \\
&\left.
\frac{f_{\mathbf{x_{p}},\mathbf{p+q}}(\mathbf{y_pz_p-z_py_p})
+f_{\mathbf{y_{p}},\mathbf{p+q}}(\mathbf{z_px_p-x_pz_p})}
{E_{e,\mathbf{p}+\mathbf{q}}+E_{l,\mathbf{p}}-\hbar\omega_q}\right],
\end{align}
where $f_{{\mathbf e}_i,{\mathbf p}}\equiv{\mathbf s}_{\mathbf p}\cdot
{\mathbf e}_i$. Using the anti-symmetric tensor
$\mathscr{E} \equiv \epsilon_{ijk}\mathbf{e}_i\mathbf{e}_j\mathbf{e}_k$ which is
invariant under orthogonal coordinate transformation, we can express
$\mathbf{x_py_p-y_px_p}=\mathbf{z_p}\cdot\mathscr{E}$, $\mathbf{y_pz_p-z_py_p}=\mathbf{x_p}\cdot\mathscr{E}$ and $\mathbf{z_px_p-x_pz_p}=\mathbf{y_p}\cdot\mathscr{E}$, whereby the terms associated with the electron spin polarization
form an anti-symmetric tensors. Noticing that the contraction between the
anti-symmetric and the symmetric tensors associated with
$I_x$ and $I_y$ must vanish, and also that the effective Hamiltonian must be real, we have
\begin{subequations}
\begin{align}
{\mathcal H}^{\rm HL}_{\rm eff} = & -\left|d_{\mathrm{cv}}\right|^2
I_z\left(\mathbf{xy}-\mathbf{yx}\right)
:\sum_{\mathbf{p}}\left[\frac{{\bf s}_{\bf p+q}\cdot\mathbf{z_pz_p}\cdot\mathscr{E}}{E_{e{{\mathbf q}+{\mathbf
p}}}+E_{h\mathbf{p}}-\hbar\omega_q}
-\frac{{\bf s}_{\bf p+q}\cdot\mathbf{z_pz_p}\cdot\mathscr{E}}{E_{e{{\mathbf q}+{\mathbf
p}}}+E_{l\mathbf{p}}-\hbar\omega_q}
 +\frac{2}{3}\frac{\mathscr{E}}
{E_{e{{\mathbf q}+{\mathbf
p}}}+E_{l\mathbf{p}}-\hbar\omega_q}\right]
\\
= & 2\left|d_{\text{
cv}}\right|^2I_z\sum\limits_{\mathbf{p}}\left[
\left(\frac{\mathbf{s_{p+q}\cdot
z_pz_p}\cdot\mathbf{z}}{E_{e{{\mathbf q}+{\mathbf
p}}}+E_{h\mathbf{p}}-\hbar\omega_q} -(E_h\rightarrow E_l)\right)+\frac{2}{3}\frac{\mathbf{s_{p+q}\cdot z}}{E_{e{{\mathbf
q}+{\mathbf p}}}+E_{l\mathbf{p}}-\hbar\omega_q}\right].\label{detail_effective_1}
\end{align}
\end{subequations}
By expanding to the first order of $\mathbf{q}$, we have $E_{e{\mathbf q}+{\mathbf p}}\approx
E_{e{\mathbf p}}+{\mathbf q}\cdot{\nabla}_{\mathbf
p}E_{e{\mathbf p}}$, and
$\mathbf{s_{p+q}}\approx \mathbf{s_p} +{\mathbf q}\cdot{
\nabla}_{\mathbf p}\mathbf{s_p}$. By using $\nabla_{\bf
p}\mathbf{z_p}=\nabla_{\bf
p}(\mathbf{p}/p)=I^{(2)}/p-\mathbf{pp}/{p^3}=(\mathbf{x_px_p+y_py_p})/{p}$,
and $\nabla_{\bf p}(\mathbf{z_pz_p}) =
(\mathbf{x_pz_px_p+y_pz_py_p+x_px_pz_p+y_py_pz_p})/{p}$, we obtain the
effective Hamiltonian as
\begin{align}
\mathcal{H}^{\text{HL}}_{\rm eff} = & 2\left|d_{\rm
cv}\right|^2I_z\sum\limits_{\mathbf{p}}\left[\left(\mathbf{s_p\cdot
z_pz_p\cdot
z}\right)\left(\frac{1}{\Delta_h}-\frac{1}{\Delta_l}\right)+\frac{2}{3}\mathbf{s_p\cdot
z}\frac{1}{\Delta_l}\right]-\frac{2}{e}\left|d_{\rm
cv}\right|^2qI_z\sum\limits_{\mathbf{p}}\mathrm{Tr}\left[\mathbb{J}_{\mathbf{p}}\right]
\left(\frac{1}{2\Delta_hE_F}-\frac{1}{2\Delta_lE_F}\right)
\nonumber
\\
&+\frac{2}{e}\left|d_{\rm
cv}\right|^2qI_z\mathbf{zz}:\sum\limits_{\mathbf{p}}\left[\mathbf{z_pz_p}\cdot\mathbb{J}_{\mathbf{p}}
\Bigg(\frac{m_e}{m_h\Delta_h^2}-\frac{m_e}{m_l\Delta_l^2}+\frac{1}{\Delta_hE_F}-\frac{1}{\Delta_lE_F}\Bigg)
+\mathbb{J}_{\mathbf{p}}
\Bigg(\frac{2}{3}\frac{m_e}{m_l\Delta_l^2}
+\frac{1}{2\Delta_lE_F}-\frac{1}{2\Delta_hE_F}\Bigg)\right],\label{HLcontribution}
\end{align}
\end{widetext}
where $E_F$ is the Fermi energy, $\Delta_{h/l}$ is the light detuning from the HH/LH band to the Fermi level, respectively~[see Fig. 1(a)]. The first term in Eq.~(\ref{HLcontribution}) results from the spin polarization, while the other terms result from a spin current. When neglecting the HH-LH splitting by letting $\Delta_h=\Delta_l$ and $m_h=m_l$, Eq.~(\ref{HLcontribution}) is reduced to a expression similar to Eq.~(\ref{SO}) but with a minus sign($m_t\rightarrow m_h$, $\Delta_t\rightarrow\Delta_h$). This reduction confirms that the effective Hamiltonian from the HH/LH-CB transitions can be derived as easy as that from the SO-CB transitions if the spin quantization direction in HH/LH band can be chosen arbitrarily. Moreover, if there were no spin-orbit coupling in the valence bands, i.e., the HH, LH, and SO bands had the same effective mass and the same band-edge energy, the coupling between a spin current and a light would vanish.

Finally, once the spin distribution is specifically given, the total effective Hamiltonian will be determined. We assume that the electron spin distribution around Fermi wavevector $k_F$ deviate only slightly from the equilibrium distribution. More specifically, we suppose the spin distribution has the form
\begin{equation}
{\bf s_p} = \mathbf{N}_0 + {\mathbf
N}_1f(p)\cos\theta_{\mathbf p}, \label{distribution}
\end{equation}
where $\theta_{\mathbf p}$ is the angle between the momentum ${\mathbf p}$ and
the current direction ${\mathbf Z}$. Such a distribution is the usual case for weak currents.
A straightforward integration over the momentum space gives [see Appendix \ref{appendix_Coordinate_basis}]
\begin{subequations}
\begin{align}
\mathcal{H}^{(0)}_{\rm eff} &= \zeta_0I_z\mathbf{z\cdot S},\label{effective_0}\\
\mathcal{H}^{(1)}_{\rm eff} &=
\zeta_1qI_zJ_Z+\zeta_2qI_z\mathbf{z}\cdot\mathbb{J}\cdot\mathbf{z},\label{effective_1}
\end{align}
\end{subequations}
with the coupling constants
\begin{subequations}\label{couplingconstatnt}
\begin{align}
\zeta_0 \equiv & \frac{2}{3}|d_{\rm cv}|^2\left(\frac{1}{\Delta_h}+\frac{1}{\Delta_l}-\frac{2}{\Delta_{t}}\right),
\\
\zeta_1 \equiv & \frac{|d_{\rm cv}|^2}{e}\Bigg(\frac{2m_e}{5m_h\Delta_h^2}-\frac{2m_e}{5m_l\Delta_l^2}-\frac{3}{5\Delta_hE_F}+\frac{3}{5\Delta_lE_F}\Bigg),
\\
\zeta_2 \equiv & \frac{|d_{\rm cv}|^2}{e}\Bigg(\frac{4m_e}{5m_h\Delta_h^2}+\frac{8m_e}{15m_l\Delta_l^2}-\frac{4m_e}{3m_{t}\Delta_{t}^2}-\frac{1}{5\Delta_hE_F}
+\frac{1}{5\Delta_lE_F}\Bigg).
\end{align}
\end{subequations}
For a spin distribution different from Eq.~(\ref{distribution}), as can be seen in Sec. IV D, the
coupling constants shown above will only be
changed quantitatively, which further confirms the symmetry analysis in Sec.
II [see Eq.~(\ref{Eq_symmetry0})].

\subsection{Second-order nonlinear optical effects}
The linear optical effect of spin currents is weak since the
photon current involves the small light momentum. If we replace the light momentum by another optical field, the coupling can be greatly enhanced by a factor of $\mathbf{F}_2\cdot\mathbf{v}/\mathbf{q}\cdot\mathbf{v}$. As shown in Sec.~\ref{sec_second}, the second-order nonlinear optical effects of spin currents is rooted in their unique physical nature and spatial inversion-symmetry breaking. Specially, noticing that a longitudinal spin
current, is a chiral quantity, we envisaged that it could be
probed by the chiral sum-frequency optical (SFG) spectroscopy which was recently developed to detect molecular chirality.~\cite{belkin2000prl,naji2006jacs,wang2005pnas} If otherwise measured in linear optics, the
effect of the chirality relies on the small magnetic moment of the molecules, and in turn on the small wave vector of the probe light, similar to the case of linear optical effects of spin currents.~\cite{wang2008prl}

\subsubsection{Physical picture}
\begin{figure}[b]
\includegraphics[width=2.6in,clip=true]{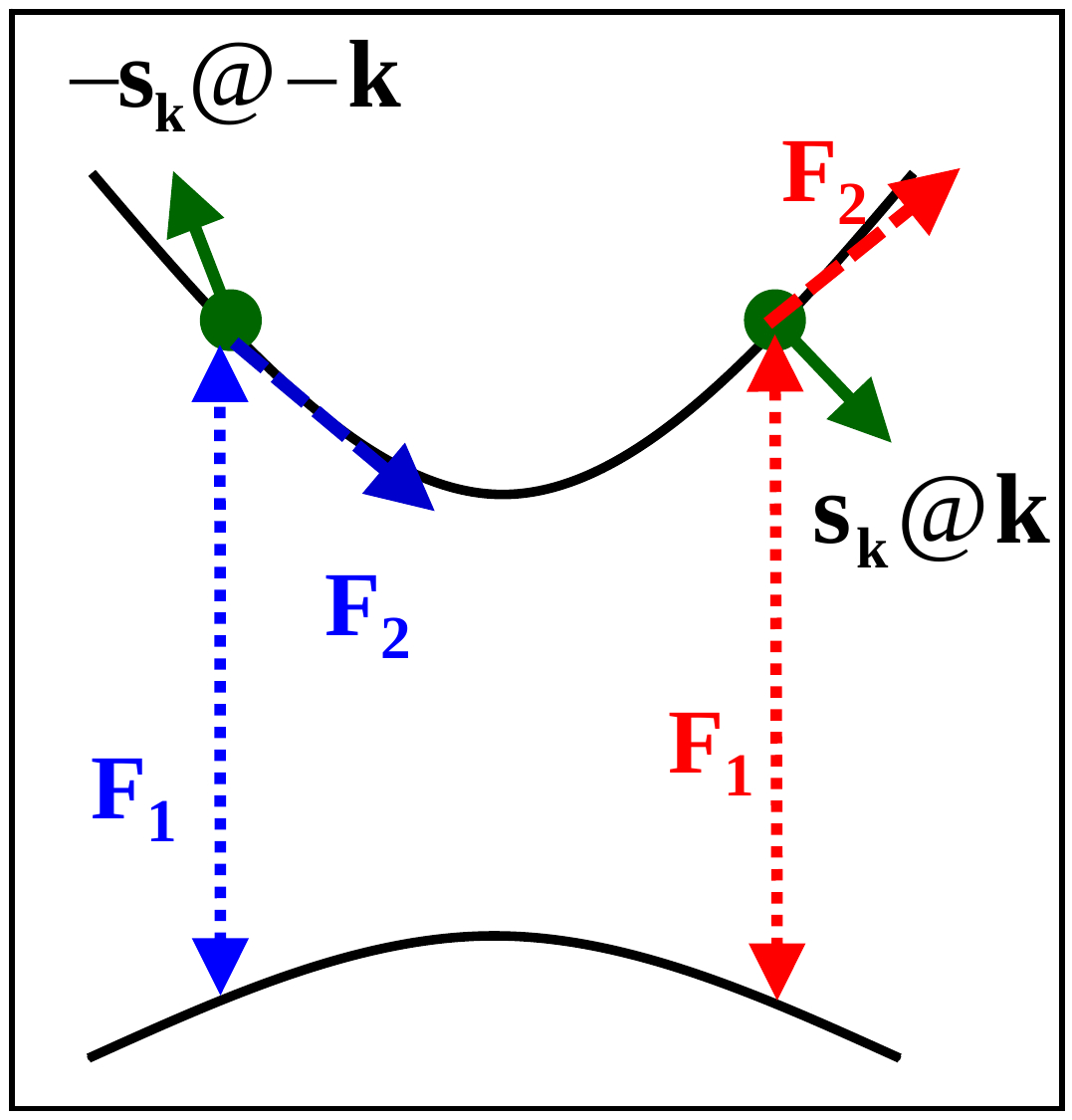}
\caption{(color online) Physical picture for the microscopic mechanism of the second-order nonlinear optical effects of a pure spin current. The second light $\mathbf{F}_2$ will accelerate the electrons (or holes).}
\label{fig3}
\end{figure}
The nonlinear coupling between a spin current and light contains three processes: one virtual interband transition creating an electron-hole pair, one intraband transition accelerating the electron or the hole, and one virtual transition inducing the combination. The physical picture for the microscopic mechanism of the second-order nonlinear optical effects of spin currents is similar to the linear optical effect.
A spin will induce a Faraday rotation $\mathbf{P}^{(1)} \propto \mathbf{F}\times\mathbf{s_k}/(\omega-E_{\mathbf{k}})$. The Faraday rotations due to the pair of spins of $\mathbf{s_k}$ (at momentum $\mathbf{k}$) and $-\mathbf{s_k}$ (at momentum $-\mathbf{k}$) cancel each other in the vertical optical transition. Instead of considering the small light-momentum in the linear optical effect, we add another optical field $\mathbf{F}_2$. The spin will experience an intraband acceleration by this optical field and the transition energy will be changed to $E_{\pm\mathbf{k}}\rightarrow E_{\pm\mathbf{k}}\pm\int e\mathbf{v_k}\cdot\mathbf{F}_2e^{-i\omega_2t}dt$~[Fig.~\ref{fig3}]. The physical meaning of $e\mathbf{v_k}\cdot\mathbf{F}_2$ is clear that is the power done by the
field to the electron. Therefore, $\pm\mathbf{s_k}$ will induce different Faraday rotation due to opposite energy modification
\begin{align}
\mathbf{P}^{(2)}&\propto\mathbf{F}_1\times\mathbf{s_kev_k}\cdot\mathbf{F}_2/[(\omega_1+\omega_2-E_{\mathbf{k}})(\omega_1-E_{\mathbf{k}})\omega_2].
\end{align}
This gives the second-order nonlinear optical effects of spin currents.

\subsubsection{Microscopic calculation}
The second-order nonlinear susceptibility can be obtained straightforwardly through the standard perturbation method as shown below. Here we take the SFG as an example of the second-order nonlinear optical effects of spin currents.

The dipole density operator for the intraband transition reads\cite{sipe2000prb}
\begin{widetext}
\begin{align}
\hat{\mathbf{P}}_{\text{intra}}(\mathbf{r})  = & ie\sum\limits_{\mathbf{k,p}}\Bigg[\sum\limits_{\mu,\mu'=\pm}
\hat{e}^{\dagger}_{\mu^{\prime},{\mathbf p}}\hat{e}_{\mu,{\mathbf
k}}\langle\mu'|_\mathbf{p}\mu\rangle_\mathbf{k}+\sum\limits_{j',m';j,m}\hat{V}^{\dagger}_{j',m';\mathbf{p}}
\hat{V}_{j,m;\mathbf{k}}\langle
j',m'|_\mathbf{p}j,m\rangle_\mathbf{k}\Bigg] \nabla_{{\mathbf k}}
e^{i\mathbf{p\cdot r}-i\mathbf{k\cdot r}}.
\end{align}
With the input optical field consisting of several frequency components $\mathbf{F}(\mathbf{r},t)=\sum\nolimits_{j=1,2}\mathbf{F}_je^{-i\omega_jt}+\text{c.c.}$,
the light-matter interaction Hamiltonian is
$\hat{H}_1(t) = -\int \hat{{\mathbf P}}({\mathbf r})\cdot{\mathbf F}({\mathbf r},t)d\mathbf{r}$, where $\hat{{\mathbf P}}({\mathbf r})=\hat{{\mathbf P}}({\mathbf r})_{\text{inter}}+\hat{{\mathbf P}}({\mathbf r})_{\text{intra}}$. Explicitly, we can write
\begin{align}
\hat{H}_1(t) = & -\left(\hat{\mathbf D}+\hat{\mathbf
D}^{\dagger}+\hat{\mathbf d}\right)\cdot \Bigg(\sum_{j=1,2}
\mathbf{F}_je^{-i\omega_jt}+{\rm c.c.}\Bigg),
\end{align}
with
\begin{subequations}
\begin{align}
\hat{\mathbf{D}} \equiv &
-{d_{\mathrm{cv}}^*}\sum\limits_{\mu,\mathbf{k}} \left(
 \mathbf{n}_{\bar{\mu},{{\mathbf k}}}\hat{h}_{\bar{\mu},-{\mathbf k}}\hat{e}_{\mu,{\mathbf k}}
+(1/\sqrt{3})\mathbf{n}_{{\mu},{{\mathbf
k}}}\hat{l}_{\mu,-{\mathbf k}}\hat{e}_{{\mu},{\mathbf k}}
-\sqrt{2/3}{\mathbf z}_{{\mathbf
k}}\hat{l}_{\bar{\mu},-{\mathbf k}}\hat{e}_{\mu,{\mathbf k}}
-\mu\sqrt{2/3}\mathbf{n}_{{\mu},{{\mathbf
k}}}\hat{t}_{\mu,-{\mathbf k}}\hat{e}_{{\mu},{\mathbf k}}
+(\mu/\sqrt{3}){\mathbf z}_{{\mathbf
k}}\hat{t}_{\bar{\mu},-{\mathbf k}}\hat{e}_{\mu,{\mathbf k}}
\right),
\\
\hat{\mathbf{d}} \equiv &
ie\sum\limits_{\mathbf{k,p}}\Bigg(\sum\limits_{\mu=\pm}\hat{e}^{\dagger}_{{\mu},{\mathbf
p}}\hat{e}_{\mu,{\mathbf
k}}+\sum\limits_{j,m}\hat{V}^{\dagger}_{j,m;\mathbf{p}}
\hat{V}_{j,m;\mathbf{k}}\Bigg)\nabla_{{\mathbf
k}}\delta_{\mathbf{p,k}}
-ie\sum\limits_{\mathbf{k}}\Bigg(\sum\limits_{\mu,\mu'=\pm}
\hat{e}^{\dag}_{\mu',\mathbf{k}}\hat{e}_{\mu,\mathbf{k}}
\langle\mu'|_{\mathbf{k}}\nabla_{\mathbf{k}}|\mu\rangle_{\mathbf{k}}
+\sum\limits_{j,m,m'}\hat{V}^{\dag}_{j,m';\mathbf{k}}\hat{V}_{j,m;\mathbf{k}}
\langle
j,m'|_{\mathbf{k}}\nabla_{\mathbf{k}}|j,m\rangle_{\mathbf{k}}\Bigg),\label{berry}
\end{align}
\end{subequations}
\end{widetext}
denoting the inter- and intra-band polarization operators, respectively. $\hat{D}$ and $\hat{D}^{\dag}$ are the positive- and negative-frequency components of the inter-band polarization operator, respectively. The first part of the intra-band polarization is the usual acceleration term. The second part, which has the form of non-Abelian Berry connections (similar to vector potentials), accounts for the variation of the spin quantization direction with acceleration of an electron. It is necessary to include the Berry connection term
for the gauge-invariance of the intra-band polarization. The explicit form of the Berry connection term depends on the choice of
the local coordinate ($\mathbf{x_p,y_p,z_p}$) at momentum $\bf p$. In Appendix \ref{appendix_Berry} we present an example for the Berry connection in a specific convention.

We adopt the interaction picture for calculating the SFG. The second-order polarization response
obtained by the standard perturbation theory is
\begin{align}
\mathbf{P}^{(2)}(t) = -\int^t_{-\infty} dt'\int^{t'}_{-\infty}
dt^{''}\mathrm{Tr}\Big{[}\hat{\tilde{\mathbf{D}}}(t)\Big[\hat{\tilde{H}}_1(t^{\prime}),\Big[\hat{\tilde{H}}_1(t^{\prime\prime}),\hat{\rho}_0\Big]\Big]\Big{]},
\nonumber
\end{align}
where $\hat{\tilde{\mathbf{D}}}(t)$, $\hat{\tilde{H}}_1$ are
operators in the interaction picture. We consider the case that (1) the sum frequency $\omega=\omega_1+\omega_2$ is near resonant
with the band-edge, so the positive-frequency component $\hat{\tilde{\mathbf{D}}}(t)$ dominates the optical process; (2) the intra-band dipole moment must be considered for the contribution by the spin current; and (3) no holes exist in the initial system, so the inter-band excitation has to be involved (caused by $\hat{D}^{\dag}$). With all these considerations taken into account, the second-order response of interest is
\begin{widetext}
\begin{subequations}\label{SFG}
\begin{align}
\mathbf{P}^{(2)}(t) =& - \int^t
dt^{\prime}\int^{t^{\prime}}dt^{\prime\prime}e^{-i\omega_2t^{\prime}-i\omega_1t^{\prime\prime}}\mathrm{Tr}
\left(\hat{\tilde{\mathbf{D}}}(t)\mathbf{F}_2\cdot\hat{\tilde{\mathbf{D}}}^{\dag}(t^{\prime})
\left[\mathbf{F}_1\cdot\hat{\tilde{\mathbf{d}}}(t^{\prime\prime}),\hat{\rho}_0\right]\right)
+ \Big\{{\mathbf F}_1, \omega_1 \leftrightarrow {\mathbf F}_2,
\omega_2 \Big\}
\label{SFG1}\\
& -\int^t
dt^{\prime}\int^{t^{\prime}}dt^{\prime\prime}e^{-i\omega_2t^{\prime\prime}-i\omega_1t^{\prime}}\mathrm{Tr}
\left(\hat{\tilde{\mathbf{D}}}(t)\left[\mathbf{F}_1\cdot
\hat{\tilde{\mathbf{d}}}(t^{\prime}),
\mathbf{F}_2\cdot\hat{\tilde{\mathbf{D}}}^{\dag}(t^{\prime\prime})\hat{\rho}_0\right]\right)
+ \Big\{{\mathbf F}_1, \omega_1 \leftrightarrow {\mathbf F}_2,
\omega_2 \Big\}.\label{SFG2}
\end{align}
\end{subequations}
\end{widetext}
The physical meaning of Eq.~(\ref{SFG}) is clear: Eq.~(\ref{SFG1}) corresponds to the driving of the electron population (at $t''$) followed by inter-band excitation (at $t'$) and emission (at $t$); Eq.~(\ref{SFG2}) corresponds to the process in which an electron-hole pair (created at $t''$) is driven by an external field (at $t'$) till its emission (at $t$).

When the HH-LH splitting is neglected, we have a simple microscopic calculation as discussed in Ref.~\onlinecite{wang2010prl}, in which the spin quantization for valence band states and the selection rule for interband transitions are independent of its momentum. Beyond such an approximation, the calculation of $\mathbf{P}^{(2)}$ through Eq.~(\ref{SFG}) is lengthy, but only quantitatively modifies
the results. So we will only list the result in the Appendix \ref{appendix_nonlinear}, and the details are shown in the Supplementary Information.

\section{Discussions and Numerics}
\label{Sec_discussion}
\subsection{Faraday rotation of a spin current and spin polarization}
The Faraday rotation angle is expressed as
\begin{equation}
\theta_{\rm F}={\omega_q
l}(\chi_{++}-\chi_{--})/\left(4nc\right),
\end{equation}
where $l$ is the light propagation distance, $n$ is the material
refractive index, and $c$ is the light velocity in vacuum [Appendix \ref{appendix_Faraday}].

\begin{figure}[htbp]
\begin{center}
\includegraphics[width=3.0in,clip=true]{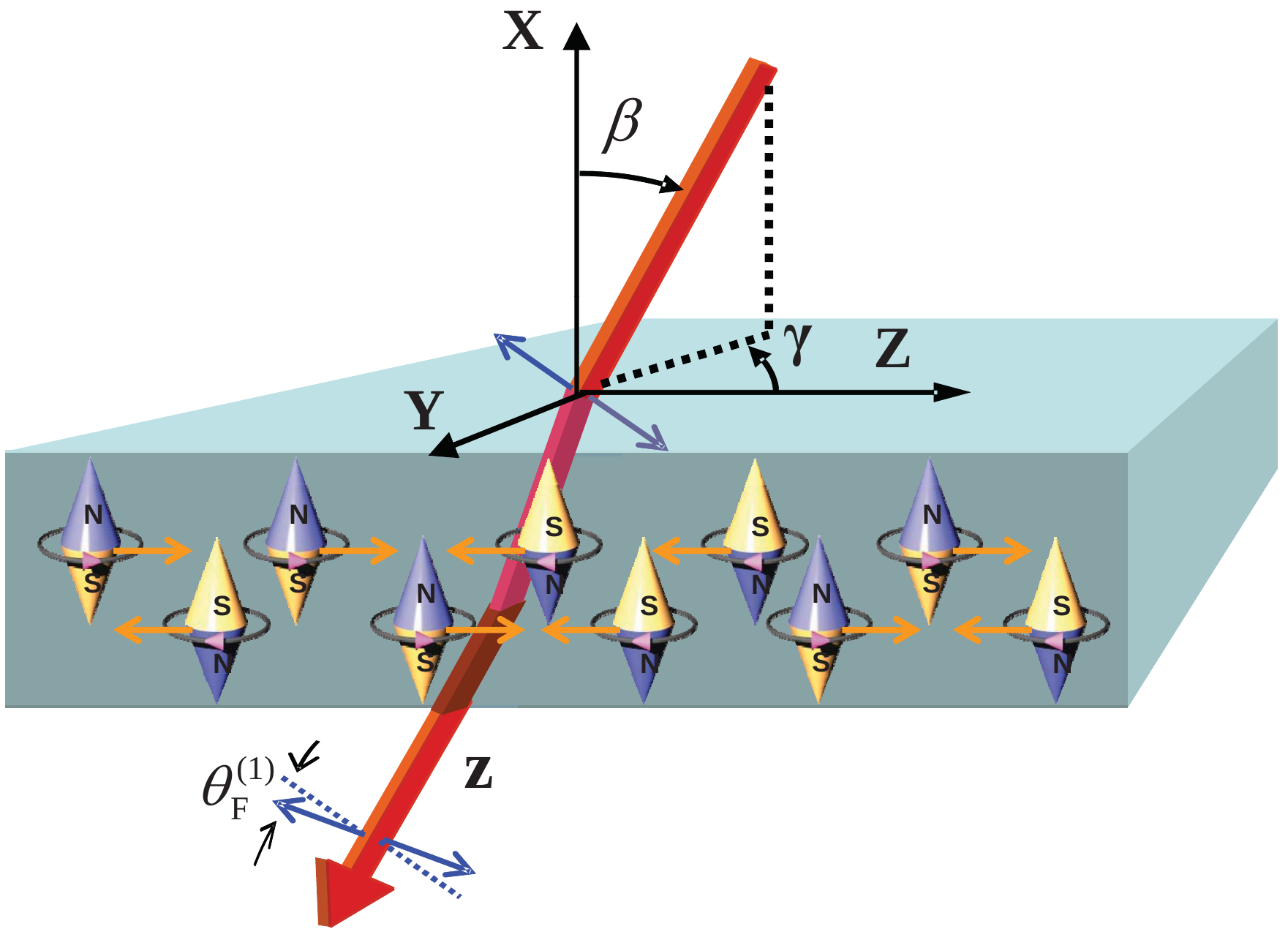}
\end{center}
\caption{ (color online) The geometry for measuring a spin current, in which the spin current is along $Z$-direction and the red arrow denotes the propagation direction of the light beam.} \label{fig4}
\end{figure}

\begin{figure}[htbp]
\begin{center}
\includegraphics[width=3.4in, clip=true]{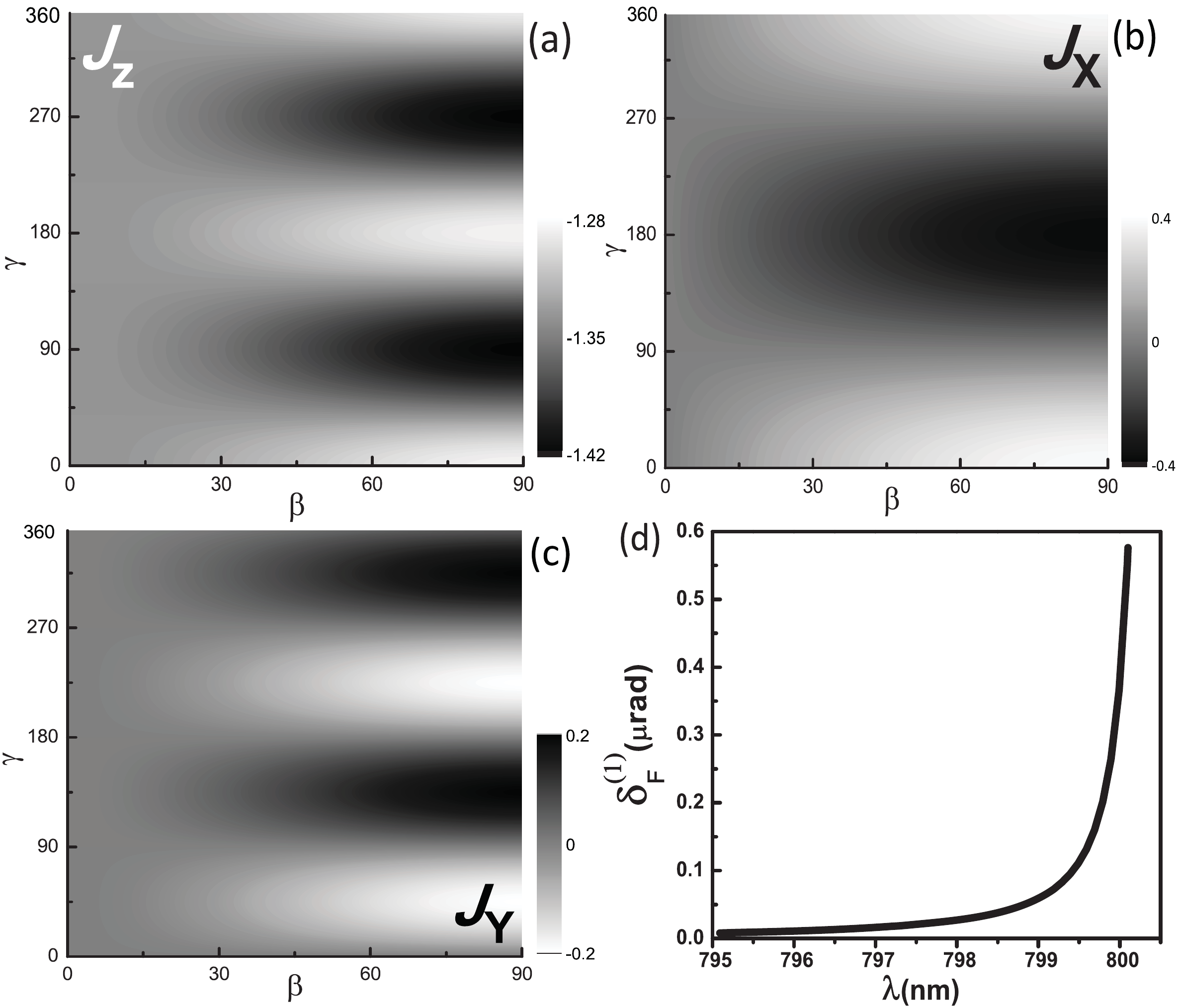}
\end{center}
\caption{(a)-(c) The Faraday rotation
amplitude of spin current components $J_Z$, $J_Y$
and $J_X$ as functions of the incident angles of the light beam. (d) The dependence of $\delta^{(1)}_{\rm F}$ on the
light wavelength $\lambda$. Parameters are chosen similar to those in Ref.~\onlinecite{kato2004science}: $E_g=$1519~meV, $E_{\rm SO}=$341~meV, the doping concentration is $3\times10^{16}$~cm$^{-3}$, the effective mass (in units of free electron mass) of the HH, LH, SO, and conduction bands is in turn 0.45, 0.082, 0.15, and 0.067, the dipole $d_{\mathrm{cv}}=6.7$~$e$\AA, $n=3.0$, $L=2.0$~$\mu$m, $E_F=5.3$~meV, and $J_X=J_Y=J_Z=20$~nA$\mu$m$^{-2}$.} \label{fig5}
\end{figure}

\emph{Pure spin current}. For a spin current configuration as shown in Fig.~\ref{fig4}, where a light comes in with a zenith angle $\beta$ and an azimuth angle $\gamma$, the Faraday rotation angle due to different components of $\mathbf{JZ}$ is
\begin{subequations}
\begin{align}
\theta^{(1)}_{\mathrm{F}}(J_X) &= \delta^{(1)}_{\rm F} J_X\zeta_2\sin\beta\cos\gamma,
\\
\theta^{(1)}_{\mathrm{F}}(J_Y) &= -\delta^{(1)}_{\rm F} J_Y\zeta_2\sin^2\beta\sin\gamma\cos\gamma(n^2-\sin^2\beta)^{-1/2},
\\
\theta^{(1)}_{\mathrm{F}}(J_Z) &= \delta^{(1)}_{\rm F} J_Z(\zeta_1n^2+\zeta_2\sin^2\beta\cos^2\gamma/n)(n^2-\sin^2\beta)^{-1/2},
\end{align}
\end{subequations}
where $\delta^{(1)}_{\rm F}={\pi^2l}/{2n\epsilon_0\lambda^2}$. The
dependence of the rotation angle on the incident angles for $J_Z$, $J_Y$
and $J_X$ components of a pure spin current are shown in turn
in Fig.~\ref{fig5} (a),(b) and (c).

\emph{Net spin polarization}. The net spin polarization also causes the Faraday rotation. With the incident light of zenith angle $\beta$, the Faraday rotation angle equals
\begin{align}
\theta^{(0)}_{\rm F}(\mathbf{S}) &= (2\pi l/8\epsilon_0n\lambda)\left(\zeta_0\mathbf{z\cdot S}\right).
\end{align}
Spin polarization has both the normal and
parallel components with respect to the sample surface
$\mathbf{S}=\mathbf{S}_\perp+\mathbf{S}_\parallel$. For
the normal component $\mathbf{S}_\perp$, the rotation is independent of $\beta$,
\begin{align}
\theta^{(0)}_{\rm F}(\mathbf{S}_\perp) &= \pi\zeta_0S_\perp L/4\epsilon_0n\lambda,
\end{align}
while for parallel component $\mathbf{S}_\parallel$,
\begin{equation}\label{S2}
\theta^{(0)}_{\rm F}(\mathbf{S}_\parallel)=(\pi\zeta_0S_\parallel L/4\epsilon_0\lambda)\sin\beta
\cos\gamma(n^2-\sin^2\beta)^{-1/2}.
\end{equation}

In general, the angle dependence of Faraday rotation can be used to distinguish a pure spin current from a spin polarization. However, in many materials $n\gg1\geq\sin\beta$, both $
\theta^{(0)}_{\rm F}(\mathbf{S}_{\parallel})$ and $\theta_{\rm F}^{(1)}(J_X)$ have nearly the same angle dependence, which is proportional to $\sin\beta\cos\gamma$. As there is inversion symmetry difference between a pure spin current ($\mathcal{P}=-$, odd) and a spin polarization ($\mathcal{P}=+$, even), a pure spin current would have a sign flip at reflection while a spin polarization would not. Therefore, the Faraday rotation angle of a pure spin current vanishes through reflection, while the rotation angle of a spin polarization will be doubled. This difference can be used to distinguish the effect of a spin current from that of spin polarization.

For the realistic case in Ref.~\onlinecite{kato2004science}, the vanishing Faraday signal is reported in
the middle region where the spin current flows without net spin polarization. We explain it with the fact that in the experiment $\mathbf{Z\cdot z}=0$ and $J_Z=0$.\cite{wang2008prl} With the experimental configuration shown in Fig.~\ref{fig4}, the rotation angle
$\theta^{(1)}_{\rm F}(\beta,\gamma) \propto \delta^{(1)}_{\rm F} \sin\beta\cos\gamma$. The maximum Faraday rotation angle is reached when $\beta\rightarrow \pi/2$ and $\gamma\rightarrow 0$. The dependence of maximum Faraday rotation angle on the light wavelength is plotted in Fig.~\ref{fig5}(d). For the specific example shown in Fig.~\ref{fig5}(d) with light wavelength around 800~nm , the maximum Faraday rotation angle is $0.38$~$\mu$rad. Such a Faraday rotation angle, though still small, is measurable in experiments.

\subsection{Effects of valence band anisotropy}\label{anisotropy}
In the derivation above, we have neglected the anisotropy of the valence bands. Now we examine the effect of the valence bands anisotropy. The anisotropic valence band Hamiltonian takes the form
\begin{align}
H^{A}_{LK} = & \frac{1}{2m_0}\left[\left(\gamma_{1}+5\gamma_{2}/2\right)\nabla^2-2\gamma_{3}
\left(\nabla\cdot\mathbf{K}\right)^2\right.
\nonumber
\\
&\left.+2\left(\gamma_{3}-\gamma_{2}\right)\left(\nabla_{\mathrm{x}}^{2}K_{\mathrm{x}}^2+\mathrm{c.p.}\right)\right],
\end{align}
where the $(\gamma_3-\gamma_2)$ term describes the anisotropy. The anisotropy is usually small. The eigenfuctions of $H^{A}_{LK}$ are
\begin{align}
|\psi_{i}\rangle=\sum\limits_{j=\pm3/2,\pm1/2}\alpha_i^j|3/2,j\rangle,\
\ i=\pm3,\pm1,
\end{align}
where the basis states $|3/2,\pm3/2\rangle$ and $|3/2,\pm1/2\rangle$ are explicitly given in Appendix~\ref{ani},
and $\alpha_i^j$ are coefficients satisfying
$U^{\dagger}\alpha=\alpha^*$, with $U=-i\sigma_x\otimes\sigma_y$. The eigenstates $|\psi_{\pm3}\rangle$ and $|\psi_{\pm1}\rangle$ have eigenvalues
$E_h$ and $E_l$, respectively. The dipole density operator can be explicitly
written as
\begin{eqnarray}
\hat{\mathbf{P}}(\mathbf{r})&=&
-e\sum\limits_{\mu,\mathbf{k},\mathbf{p}}e^{-i\mathbf{p}\cdot\mathbf{r}+i\mathbf{k}\cdot\mathbf{r}}\Big[\hat
h_{-,-\mathbf{p}}\hat
e_{\mu,\mathbf{k}}\langle\psi_{3}|\mathbf{r}|\mu\rangle+\hat
l_{-,-\mathbf{p}}\hat
e_{\mu,\mathbf{k}}\langle\psi_{1}|\mathbf{r}|\mu\rangle \nonumber\\
&+&\hat l_{+,-\mathbf{p}} \hat
e_{\mu,\mathbf{k}}\langle\psi_{-1}|\mathbf{r}|\mu\rangle+\hat
h_{+,-\mathbf{p}}\hat
e_{\mu,\mathbf{k}}\langle\psi_{-3}|\mathbf{r}|\mu\rangle \Big] +
\text{H.c.},
\end{eqnarray}
where $|\mu\rangle=|\pm\rangle$ denotes the CB electron state with spin $\pm1/2$, and the operators $\hat{h}_{\mp}$ and $\hat{l}_{\mp}$
annihilate $|\psi_{\pm3}\rangle$ and
$|\psi_{\pm1}\rangle$, respectively. By using the fact
$\mathbf{p}/m_0=d\mathbf{r}/dt=(\mathbf{r}H_{0}-H_{0}\mathbf{r})/i$, we get
\begin{equation}
\langle\psi_i|-e\mathbf{r}|\mu\rangle=\sum\limits_{j=x,y,z}M_{i,\mathbf{p}}A_{i,\mu}^j\mathbf{j}_{\mathbf{p}},
\end{equation} 
where $M_{\pm3/\pm1,\mathbf{p}}=-ie/m_0
(E_{e,\mathbf{p}}-E_{h/l,\mathbf{p}})$. The detailed expression for
$A_{i,\nu}^j$ can be found in Appendix C. The effective Hamiltonian then reads
\begin{align}
\mathcal{H}^A_{\mathrm{eff}} = &
\mathrm{Tr}\Bigg(\hat{\rho}\sum\limits_{\sigma,\sigma^{\prime},\mathbf{p},i,\mu,\mu^{\prime}}^{\mathbf{j},\mathbf{j}^{\prime}=\mathbf{x,y,z}}
F^*_{\sigma}F_{\sigma^{\prime}}\mathbf{n}_{\sigma^{\prime}}\mathbf{n}_{\sigma}^*:
\nonumber
\\
&\left|M_{i,\mathbf{p}}\right|^2A_{i,\mu}^jA_{i,\mu'}^{j'}\mathbf{j}_{\mathbf{p}}\mathbf{j'}_{\mathbf{p}}
\frac{1-f_{\mu\mu',\mathbf{p+q}}}{E_{e,\mathbf{p+q}}+E_{i,\mathbf{p}}-\hbar\omega_{\mathbf{q}}}\Bigg).
\end{align}
The calculation is lengthy. Here we omit the
details but just give the terms with $i=3$, $\mu=\mu^{\prime}=+$ and
$i=-3$, $\mu=\mu^{\prime}=-$ explicitly, which is
proportional to
\begin{subequations}
\begin{align}
&(|A^1_{3,+}|^{2}\mathbf{x}_{\mathbf{p}}\mathbf{x}_{\mathbf{p}}
+|A^2_{3,+}|^{2}\mathbf{y}_{\mathbf{p}}\mathbf{y}_{\mathbf{p}}
+|A^3_{3,+}|^{2}\mathbf{z}_{\mathbf{p}}\mathbf{z}_{\mathbf{p}})(f_{++,\mathbf{p}}+f_{--,\mathbf{p}})\label{background}
\\
&+2i\Big[\Im(A^1_{3,+}A_{3,+}^{2*})
(\mathbf{x}_{\mathbf{p}}\mathbf{y}_{\mathbf{p}}-\mathbf{y}_{\mathbf{p}}\mathbf{x}_{\mathbf{p}})
+2i\Im(A^2_{3,+}A_{3,+}^{3*})
(\mathbf{y}_{\mathbf{p}}\mathbf{z}_{\mathbf{p}}-\mathbf{z}_{\mathbf{p}}\mathbf{y}_{\mathbf{p}})\nonumber\\
&+2i\Im(A_{3,+}^{3}A_{3,+}^{1*})
(\mathbf{z}_{\mathbf{p}}\mathbf{x}_{\mathbf{p}}-\mathbf{x}_{\mathbf{p}}\mathbf{z}_{\mathbf{p}})\Big]
f_{\mathbf{z}_\mathbf{p}}.\label{anisotropic-tensor}
\end{align}
\end{subequations}
The term (\ref{background}) is just a background. The term
(\ref{anisotropic-tensor}) is the total anisotropy-tensor, which
couples to $I_{z}$ only. This result confirms the symmetry analysis in Sec.~\ref{Sec_symmetry}.

\subsection{Second-order nonlinear optical effects}
\begin{figure}[htbp]
\begin{center}
\includegraphics[width=2.5in, bb=54 13 342 265, clip=true]{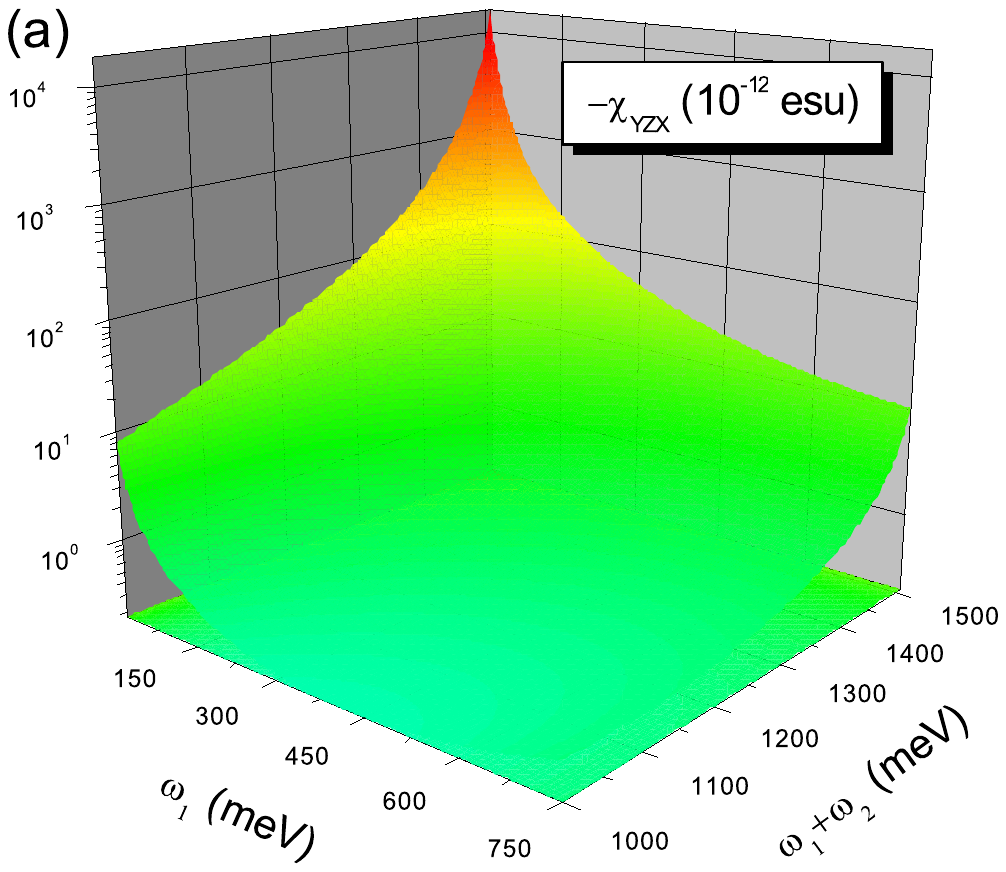}
\hfill
\includegraphics[width=2.5in, bb=54 13 342 265, clip=true]{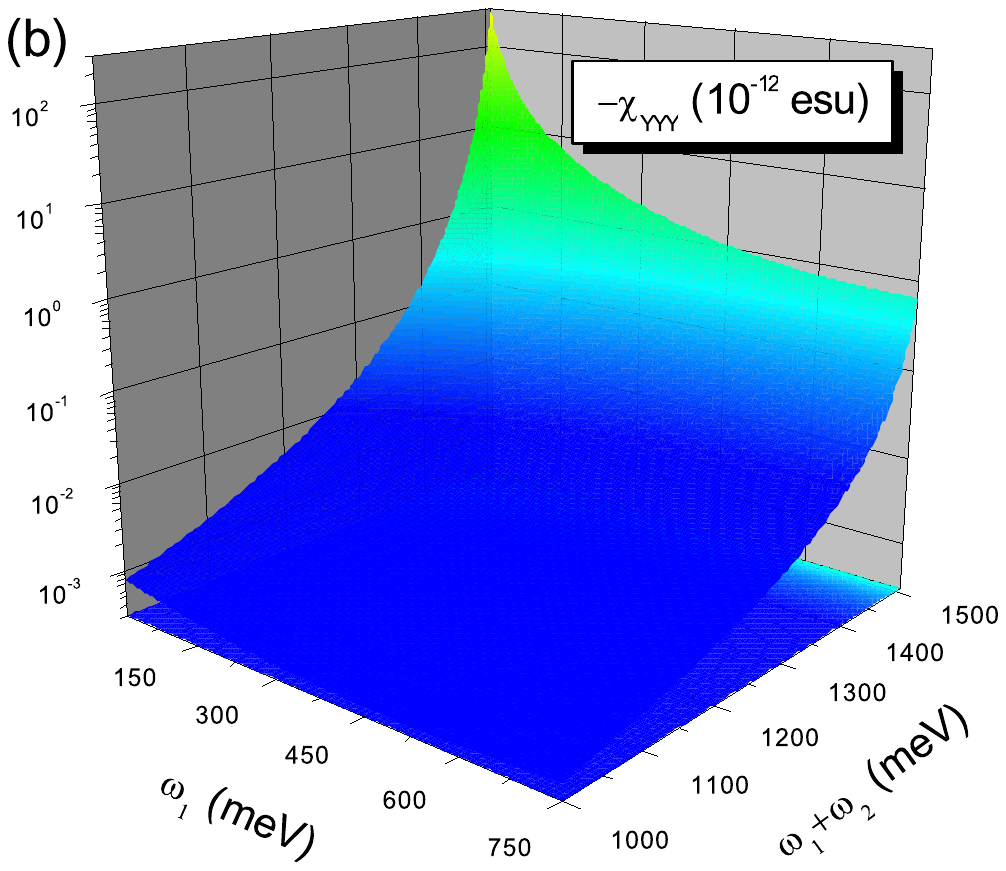}
\end{center}
\caption{Representative results of the sum frequency
susceptibility. (a) $-\chi^{(2)}_{YZX}$ due to a longitudinal spin
current, and (b) $-\chi^{(2)}_{YYY}$ due to a transverse spin current, as
functions of the optical frequencies. Parameters are chosen
similar to those in Ref.~\onlinecite{kato2004science} (same as in Fig.~\ref{fig5}). The dielectric constant
$\epsilon_r=10.6$, and the spin current $J_X=J_Z=20$~nA/$\mu$m$^{2}$.}
\label{fig6}
\end{figure}

The independent parameters of the susceptibility of spin current in a bulk GaAs in Eqs.~(\ref{chi2Jz}) and (\ref{chi2Jx}) are listed in Appendix~\ref{appendix_nonlinear}. For the sake of simplicity, we neglected the anisotropy of the valence bands. We also neglected the
Coulomb interaction, since it is largely screened in the $n$-doped material. These approximations, according to the symmetry analysis, would only quantitatively modify the results. The bulk inversion asymmetry would cause a background second-order susceptibility, which is indeed strong but can be well separated from the spin-current effect by ac modulation of the current and the phase-locking detection technique. Two representative
results of the calculated susceptibility spectra are shown in Fig.~\ref{fig6}. The other terms of the susceptibility tensor (not shown) have similar frequency dependence and comparable amplitudes. As a specific example, a transverse
spin current $20$nA/$\mu$m$^{2}$ has a susceptibility $-\chi^{(2)}_{YYY}\approx4.8\times10^{-12}$esu (or $0.2\times10^{-14}$ m/V in SI units) for input frequencies $\omega_1$=100~meV and $\omega_2=1400$~meV
or $0.25\times10^{-12}$~esu for $\omega_1=\omega_2=750$~meV (corresponding to the second harmonics generation).

The SFG of spin current can be straightforwardly extended to other second-order optical spectroscopy such as
difference-frequency and three-wave mixing.\cite{YRShen}

\section{Conclusions}
\label{Sec_conclusion}

In summary, with the systematic symmetry analysis in general and the microscopic calculation under realistic conditions,
we have shown that a pure spin current has a
measurable circular birefringence effect and a sizable sum-frequency susceptibility. With
universality of the method guaranteed by the symmetry
principle and without requirements of special structure design and fabrication, the linear and nonlinear
optical spectroscopy can be applied to study a wide range of spin-related quantum phenomena such as in topological
insulators~\cite{fu2007prl,bernevig2006science,konig2007science,hsieh2008nature,qi2010phystoday,hasan2010rmp}. A wealth of physics connecting spins and photons and technologies synthesizing spintronics
and photonics may be explored.

\begin{acknowledgments}
This work was supported by Hong Kong RGC/GRF CUHK 401011, the NSFC Grant Nos.
10774086, 10574076 and the Basic Research Program of China Grant No.
2006CB921500.
\end{acknowledgments}

\begin{appendix}
\section{Coordinate basis}
\label{appendix_Coordinate_basis}
We choose a global coordinate system ($\bf X, Y, Z$) and
define the local coordinates as
\begin{subequations}\label{coordinate}
\begin{align}
\mathbf{\hat{\theta}_p}=&\mathbf{x_p}=\mathbf{X}\cos\theta_{\mathbf{p}}\cos\phi_{\mathbf{p}}
+\mathbf{Y}\cos\theta_{\mathbf{p}}\sin\phi_{\mathbf{p}}-\mathbf{Z}\sin\theta_{\mathbf{p}},
\\
\mathbf{\hat{\phi}_p}=&\mathbf{y_p}=-\mathbf{X}\sin\phi_{\mathbf{p}}
+\mathbf{Y}\cos\phi_{\mathbf{p}},
\\
\mathbf{\hat{p}}=&\mathbf{z_p}=\mathbf{X}\sin\theta_{\mathbf{p}}\cos\phi_{\mathbf{p}}
+\mathbf{Y}\sin\theta_{\mathbf{p}}\sin\phi_{\mathbf{p}}+\mathbf{Z}\cos\theta_{\mathbf{p}}.
\end{align}
\end{subequations}
The angle average of the tensor
\begin{equation}
\overline{\mathbf{z_pz_p}} \equiv \frac{1}{4\pi}\int
\mathbf{z_pz_p}d\Omega = \frac{1}{3}I^{(2)}.
\end{equation}
And the angle average
\begin{align}
&\overline{\mathbf{Z\cdot z_pz_pz_pz_p}}\nonumber
\\
&\equiv \frac{1}{4\pi}\int
\mathbf{\mathbf{Z\cdot z_pz_pz_pz_p}}d\Omega
\nonumber
\\
&=+\overline{\cos^2\theta_{\bf p}\sin^2\theta_{\bf p}\cos^2\phi_{\bf p}}\left(\mathbf{XXZ+XZX+ZXX}\right)\nonumber
\\
&+\overline{\cos^2\theta_{\bf p}\sin^2\theta_{\bf p}\sin^2\phi_{\bf p}}\left(\mathbf{YYZ+YZY+ZYY}\right)\nonumber
\\
&+\overline{\cos^4\theta_{\bf p}}\mathbf{ZZZ}\nonumber
\\
&=\frac{1}{15}\left(I^{(2)}\mathbf{Z}+\mathbf{XZX+YZY+ZZZ}+\mathbf{Z}I^{(2)}\right).
\end{align}
For a spin distribution of Eq.~(\ref{distribution}), the total spin
polarization and the spin current is respectively as
\begin{subequations}
\begin{align}
\mathbf{S} =&
\sum\limits_{\mathbf{p}}\mathbf{s_p}=\sum\limits_{\mathbf{p}}\mathbf{N}_0,
\\
\mathbb{J} =&
\sum\limits_{\mathbf{p}}\mathbb{J}_{\mathbf{p}}=\frac{e}{m_e}\sum\limits_{\mathbf{p}}\mathbf{N}_1f(p)p\mathbf{Z\cdot
z_pz_p}\nonumber
\\
=&\frac{e}{m_e}\sum\limits_{\mathbf{p}}\mathbf{N}_1f(p)p\mathbf{Z}\cdot\overline{\mathbf{z_pz_p}}\nonumber
\\
=&\frac{\mathbf{N_1Z}}{3}\frac{e}{m_e}\sum\limits_{\mathbf{p}}f(p)p=\mathbf{JZ}.
\end{align}
\end{subequations}
Also, we have
\begin{eqnarray}
\sum\limits_{\mathbf{p}}\mathbf{z_pz_p}\cdot\mathbb{J}_{\mathbf{p}}
&=&
\frac{e}{m_e}\sum\limits_{\mathbf{p}}\mathbf{z_pz_p}\cdot\left(\mathbf{N}_1f(p)p\mathbf{Z}\cdot\mathbf{z_pz_p}\right)\nonumber
\\
&=&
\frac{e}{m_e}\sum\limits_{\mathbf{p}}f(p)p\overline{\mathbf{Z\cdot
z_pz_pz_pz_p}}\cdot\mathbf{N}_1\nonumber
\\
&=&\frac{1}{3}\frac{e}{m_e}\sum\limits_{\mathbf{p}}f(p)p\nonumber
\\
&&\times\frac{I^{(2)}\mathbf{Z}+\mathbf{XZX+YZY+ZZZ}+\mathbf{Z}I^{(2)}}{3}\cdot\mathbf{N}_1\nonumber
\\
&=&\frac{1}{5}\left(J_ZI^{(2)}+\mathbb{J}+\mathbb{J}^T\right).
\end{eqnarray}

\section{Berry connection}
\label{appendix_Berry}
The band edge state of CB are
\begin{align}
\left|+/-\right\rangle_{\mathbf{p}}=|S\rangle \otimes
|\uparrow/\downarrow\rangle_{\mathbf{p}},
\end{align}
with $|S\rangle$ being a periodic s-wave orbital wavefunction which is isotropic in a unit cell, and $|\uparrow/\downarrow\rangle_{\mathbf{p}}$ denoting the spin eigen state
parallel/anti-parallel to the momentum.

Similarly, the band edge states of the valence bands are
\begin{subequations}
\begin{align}
&\left|\frac{3}{2},+\frac{3}{2}\right\rangle_\mathbf{p}
 = -\frac{|X\rangle_\mathbf{p}+i|Y\rangle_\mathbf{p}}{\sqrt{2}}\otimes |\uparrow\rangle_\mathbf{p},\\
&\left|\frac{3}{2},+\frac{1}{2}\right\rangle_\mathbf{p}
 = \sqrt{\frac{2}{3}}|Z\rangle_\mathbf{p}\otimes |\uparrow\rangle_\mathbf{p}
 -\frac{|X\rangle_\mathbf{p}+i|Y\rangle_\mathbf{p}}{\sqrt{6}}\otimes |\downarrow\rangle_\mathbf{p}, \\
&\left|\frac{3}{2},-\frac{1}{2}\right\rangle_\mathbf{p}
 =\sqrt{\frac{2}{3}}|Z\rangle_\mathbf{p}\otimes |\downarrow\rangle_\mathbf{p}
 + \frac{|X\rangle_\mathbf{p}-i|Y\rangle_\mathbf{p}}{\sqrt{6}}\otimes |\uparrow\rangle_\mathbf{p}, \\
&\left|\frac{3}{2},-\frac{3}{2}\right\rangle_\mathbf{p}
 =+\frac{|X\rangle_\mathbf{p}-i|Y\rangle_\mathbf{p}}{\sqrt{2}}\otimes
 |\downarrow\rangle_\mathbf{p},
 \\
&\left|\frac{1}{2},+\frac{1}{2}\right\rangle_\mathbf{p}
 = -\frac{1}{\sqrt{3}}|Z\rangle_\mathbf{p}\otimes |\uparrow\rangle_\mathbf{p}-
 \frac{|X\rangle_\mathbf{p}+i|Y\rangle_\mathbf{p}}{\sqrt{3}}\otimes |\downarrow\rangle_\mathbf{p}, \\
&\left|\frac{1}{2},-\frac{1}{2}\right\rangle_\mathbf{p}
 =+{\frac{1}{\sqrt{3}}}|Z\rangle_\mathbf{p}\otimes |\downarrow\rangle_\mathbf{p}
 -\frac{|X\rangle_\mathbf{p}-i|Y\rangle_\mathbf{p}}{\sqrt{3}}\otimes |\uparrow\rangle_\mathbf{p}.
\end{align}
\end{subequations}
where $|X\rangle_\mathbf{p}, |Y\rangle_\mathbf{p},
|Z\rangle_\mathbf{p}$ are the $p$-type orbital parts of the Bloch
amplitudes with wave vector $\mathbf{p}$, which have the same rotation and inversion transformation
properties as the coordinate system $\bf x_p, y_p, z_p$, defined with respect to the momentum direction (i.e., $\mathbf{z_p}=\mathbf{p}/p$). The mixing of the orbital wavefunctions and the electron spin states in the total angular momentum eigen states includes the spin-orbit coupling automatically. This spin-orbit coupling is an intrinsic relativistic effect and does not reply on whether or not the material has inversion symmetry.

With the convention chosen in Eq.~(\ref{coordinate}), the transformation of the Bloch states
and spin states are as follows
\begin{subequations}
\begin{align}
|X\rangle_{\mathbf{p}}=&|X\rangle\cos\theta_{\mathbf{p}}\cos\phi_{\mathbf{p}}
+|Y\rangle\cos\theta_{\mathbf{p}}\sin\phi_{\mathbf{p}}-|Z\rangle\sin\theta_{\mathbf{p}},
\\
|Y\rangle_{\mathbf{p}}=&-|X\rangle\sin\phi_{\mathbf{p}}
+|Y\rangle\cos\phi_{\mathbf{p}},
\\
|Z\rangle_{\mathbf{p}}=&|X\rangle\sin\theta_{\mathbf{p}}\cos\phi_{\mathbf{p}}
+|Y\rangle\sin\theta_{\mathbf{p}}\sin\phi_{\mathbf{p}}+|Z\rangle\cos\theta_{\mathbf{p}},
\\
\left|\uparrow\right\rangle_{\mathbf{p}}=&+\cos\frac{\theta_{\mathbf{p}}}{2}e^{-i\phi_{\mathbf{p}}/2}
\left|\uparrow\right\rangle+\sin\frac{\theta_{\mathbf{p}}}{2}e^{+i\phi_{\mathbf{p}}/2}
\left|\downarrow\right\rangle,
\\
\left|\downarrow\right\rangle_{\mathbf{p}}=&-\sin\frac{\theta_{\mathbf{p}}}{2}e^{-i\phi_{\mathbf{p}}/2}
\left|\uparrow\right\rangle+\cos\frac{\theta_{\mathbf{p}}}{2}e^{+i\phi_{\mathbf{p}}/2}
\left|\downarrow\right\rangle.
\end{align}
\end{subequations}
$|X\rangle, |Y\rangle, |Z\rangle$ are the orbital Bloch functions which
transform as $X, Y, Z$, and $|\uparrow/\downarrow\rangle$ are the spin
Bloch function as the eigenstates of
$\boldsymbol{\sigma}\cdot\mathbf{Z}$ with eigenvalue $\pm1$. With
this convention, the Berry curvature term has a very simple form as
\begin{align}
-i\left\langle
j,m'\right|_{\mathbf{p}}\nabla_{\mathbf{p}}\left|j,m\right\rangle_{\mathbf{p}}
= & i\frac{\mathbf{n}_{\pm,\mathbf{p}}}{\sqrt{2}p}\delta_{m'\pm1,m}\sqrt{(j\pm
m)(j\mp
m+1)}
\nonumber
\\
& -\delta_{m,m'}m\frac{\cos\theta_{\mathbf{p}}}{\sin\theta_{\mathbf{p}}}\frac{\mathbf{y_p}}{p}.
\end{align}

\section{Faraday rotation angle}
\label{appendix_Faraday}
For a light with frequency $\omega_q$, the polarization density is
\begin{equation}
\mathbf{P} =
\epsilon_0\sum\limits_{\sigma,\sigma^{\prime}}\mathbf{n}_{\sigma}\chi_{\sigma,\sigma^{\prime}}F_{\sigma^{\prime}}.
\end{equation}
Then the energy density in the material is
\begin{eqnarray}
\rho_E &=& \frac{1}{2}\langle
\mathbf{D}(\mathbf{r},t)\cdot\mathbf{F}(\mathbf{r},t)\rangle +
\frac{1}{2}\langle\mathbf{B}(\mathbf{r},t)\cdot\mathbf{H}(\mathbf{r},t)\rangle\nonumber
\\
&=&\left(\epsilon_0\epsilon_r\mathbf{F}+\mathbf{P}\right)\cdot\mathbf{F}^*+\mathrm{c.c.},
\end{eqnarray}
where $\epsilon_r$ is the background dielectric constant. Thus the
linear optical susceptibility is related to the effective
Hamiltonian through
\begin{equation}
\mathcal{H}_{\rm eff} =
\epsilon_0\sum\limits_{\sigma,\sigma^{\prime}}\chi_{\sigma,\sigma^{\prime}}F^*_{\sigma}F_{\sigma^{\prime}}
+\epsilon_0\sum\limits_{\sigma,\sigma^{\prime}}\chi^{*}_{\sigma,\sigma^{\prime}}F_{\sigma}F^*_{\sigma^{\prime}}.
\end{equation}
Thus we have
\begin{equation}
\chi_{\sigma,\sigma^{\prime}}+\chi^*_{\sigma^{\prime},\sigma}=\frac{1}{\epsilon_0}\frac{\partial^2\mathcal{H}_{\rm
eff}}{\partial F^*_{\sigma}\partial F_{\sigma^{\prime}}}.
\end{equation}
The index change due to two circular polarization is respectively
\begin{equation}
\delta n_{\pm} =
\sqrt{\epsilon_r+\chi_{\pm\pm}}-\sqrt{\epsilon_r} \approx \pm\frac{1}{2}\chi_{\pm\pm}/\sqrt{\epsilon_r}=\pm\frac{1}{2}n^{-1}\chi_{++},
\end{equation}
where $n$ is the material refractive index. The phase delay within a
propagation length $l$ is then
\begin{equation}
\delta\phi_{\pm} =
\omega_qc^{-1}l\delta{n}_{\pm}=2\pi\lambda^{-1}l\delta{n}_{\pm},
\end{equation}
where $\lambda$ is the light wavelength in vacuum. For a light with linear
polarization
\begin{equation}
\mathbf{x} = \left(-\mathbf{n}_++\mathbf{n}_-\right)/\sqrt{2},
\end{equation}
after propagation of the length $l$, the polarization becomes
\begin{equation}
\left(-\mathbf{n}_+e^{-i\delta\phi_+}+\mathbf{n}_-e^{-i\delta\phi_-}\right)/\sqrt{2}
=\cos\delta\phi_+\mathbf{x}+\sin\delta\phi_+\mathbf{y}.
\end{equation}
So the Faraday rotation angle is
\begin{equation}
\theta_F=\delta\phi_+=\frac{2\pi l}{2n\lambda}\chi_{++}.
\end{equation}
For a light with incident zenith and azimuth angles $\beta$ and
$\gamma$, the angles inside the sample $\beta^{\prime}$ and
$\gamma^{\prime}$ are determined by
\begin{subequations}
\begin{eqnarray}
n\sin\beta^{\prime}&=&\sin\beta,\\
\gamma^{\prime}&=&\gamma,
\end{eqnarray}
\end{subequations}
the propagation length through a sample of thickness $L$ is
\begin{equation}
l=L/\cos\beta^{\prime}.
\end{equation}
For a pure spin current and an off-resonant probe, the
susceptibility is
\begin{equation}
\chi^{(1)}_{++}=-\chi^{(1)}_{--}=\frac{1}{4\epsilon_0}\left(\zeta_1qJ_Z+\zeta_2q\mathbf{z}\cdot\mathbb{J}\cdot\mathbf{z}\right),
\end{equation}
Thus the Faraday rotation for a spin current polarized normal to the
surface (as in Awschalom's experiment\cite{kato2004science}) is
\begin{eqnarray}
\theta^{(1)}_F=\delta\phi_+&=&\frac{2\pi
qL}{8\epsilon_0n\lambda\cos\beta^{\prime}}\zeta_2J\cos\beta^{\prime}\sin\beta^{\prime}\cos{\gamma}\nonumber
\\
&=&\frac{\pi^2\zeta_2JL}{2n\epsilon_0\lambda^2}\sin\beta\cos\gamma,
\end{eqnarray}
where $q=2\pi n/\lambda$ has been used.

\section{Anisotropic valence band effect}\label{ani}

The anisotropic Luttinger-Kohn matrix of $H^{A}_{LK}$ is
\begin{align}
H^A_{LK} &=
\begin{pmatrix}
E_{3} & P & Q & 0 \\
{P^*} & E_{1} & 0 & Q \\
{Q^*} & 0 & E_{-1} & -P \\
0 & {Q^*} & {-P^*} & E_{-3}
\end{pmatrix},\label{full}
\end{align}
where
\begin{subequations}
\begin{eqnarray}
E_{3}&=&E_{-3}=\frac{1}{2m_0}\left[(\gamma_{1}+\gamma_{2}){k^2}-3\gamma_{2}{k_{z}}^2\right],
\\
E_{1}&=&E_{-1}=\frac{1}{2m_)}\left[(\gamma_{1}-\gamma_{2}){k^2}+3\gamma_{2}{k_{z}}^2\right],
\\
P&=&-\frac{\sqrt{3}\gamma_{3}}{m_0}k_{z}(k_{x}-ik_{y}),
\\
Q&=&\frac{1}{2m_0}\left[-\sqrt{3}\gamma_{2}(k_{x}^2-k_{y}^2)+i2\sqrt{3}\gamma_{3}k_{x}k_{y}\right].
\end{eqnarray}
\end{subequations}
The eigenstates can be in general written as
\begin{equation}
|\psi_{i}\rangle=\sum\limits_{j=\pm\frac{3}{2},\pm\frac{1}{2}}\alpha_i^j|\frac{3}{2},j\rangle,\
\ i=\pm3,\pm1.
\end{equation}
By making the transformation $U^\dag HUU^\dag\alpha=U^\dag\alpha$ with
\begin{equation}
U^\dag=\left(
\begin{array}{cccc}
0 & 0 & 0 & -1 \\ 0 & 0 & 1 & 0 \\ 0 & -1 & 0 & 0 \\ 1 & 0 & 0 & 0
\end{array}\right),
\end{equation}
\begin{figure}[htbp]
\begin{center}
\includegraphics[width=2.6in,clip=true]{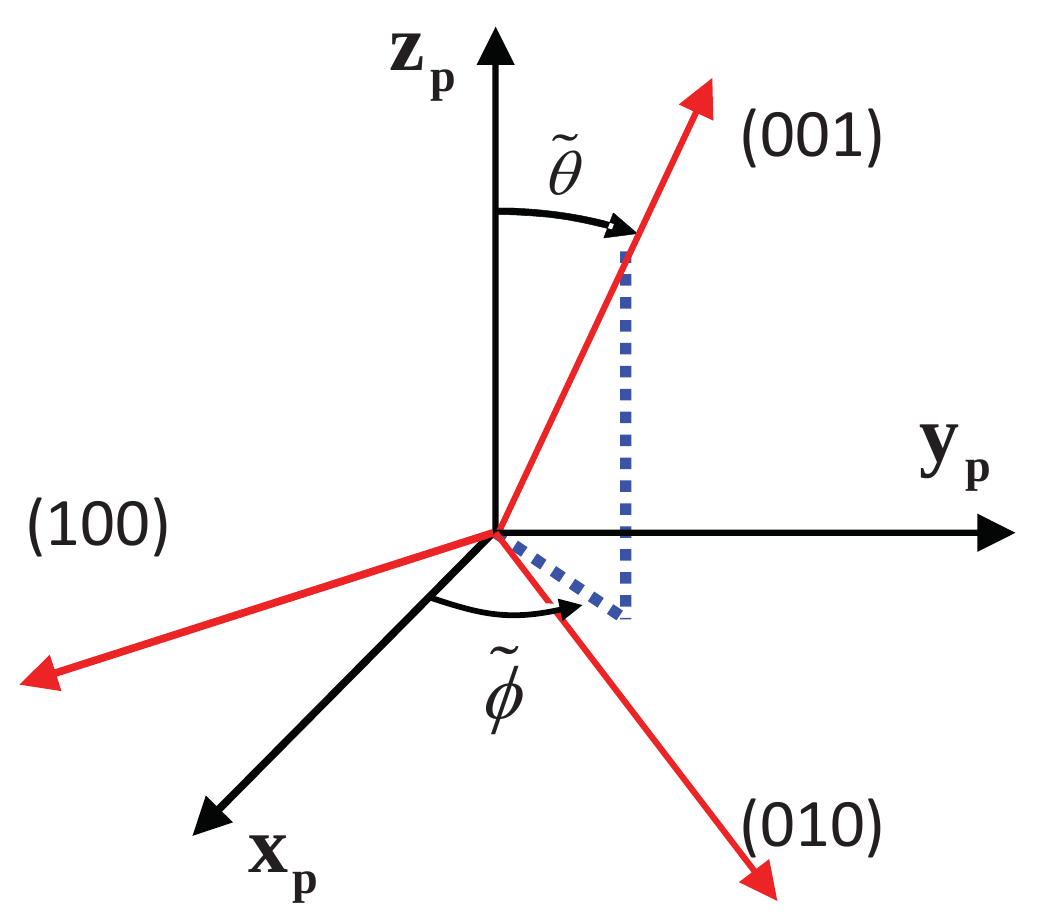}
\end{center}
\caption{ (color online) The geometry of the coordinates.} \label{fig7}
\end{figure}
we can see $U^{\dagger}HU=H^*$. Thus we get the relation
\begin{equation}\label{relation}
U^{\dagger}\alpha=\alpha^*.
\end{equation}

Without loss of generality, we take the coordinate relation
between the electron and the crystal as
\begin{eqnarray}
{(100)}&=&\sin{\tilde
\phi}\mathbf{x}_{\mathbf{p}}-\cos{\tilde
\phi}\mathbf{y}_{\mathbf{p}}\nonumber
\\
{(010)}&=&\cos{\tilde \theta}\cos{\tilde
\phi}\mathbf{x}_{\mathbf{p}}+\cos{\tilde \theta}\sin{\tilde
\phi}\mathbf{y}_{\mathbf{p}}-\sin{\tilde
\theta}\mathbf{z}_{\mathbf{p}}\nonumber
\\
{(001)}&=&\sin{\tilde \theta} \cos{\tilde
\phi}\mathbf{x}_{\mathbf{p}}+\sin{\tilde \theta}\sin{\tilde
\phi}\mathbf{y}_{\mathbf{p}}+\cos{\tilde
\theta}\mathbf{z}_{\mathbf{p}}
\end{eqnarray}
where $({001}),({010})$ and $({100})$ are
directions of the three crystal axis, and ${\tilde
\theta}$ and ${\tilde \phi}$ are the relative direction angles
between $\mathbf{x_p, y_p, z_p}$ and the $({100}),
({010}), ({001})$ axes.

\begin{widetext}
The explicit form of $A^j_{i,\mu}$ is
\begin{subequations}
\begin{align}\label{A1}
A^{x}_{i,\pm}&=\left[\mp\frac{1}{\sqrt{2}}\left(\alpha_{i}^{\pm\frac{3}{2}}\right)^*
\pm\frac{1}{\sqrt{6}}\left(\alpha_{i}^{\mp\frac{1}{2}}\right)^*\right]\sin{\tilde
\phi} +i\left[\frac{1}{\sqrt{2}}\left(\alpha_{i}^{\pm\frac{3}{2}}\right)^*
+\frac{1}{\sqrt{6}}\left(\alpha_{i}^{\mp\frac{1}{2}}\right)^*\right]\cos{\tilde
\theta} \cos{\tilde \phi}
 +\sqrt{\frac{2}{3}}\left(\alpha_{i}^{\pm\frac{1}{2}}\right)^* \sin{\tilde \theta} \cos{\tilde \phi},
 \\
A^y_{i,\pm}&=\left[\pm\frac{1}{\sqrt{2}}\left(\alpha_{i}^{\pm\frac{3}{2}}\right)^*
\mp\frac{1}{\sqrt{6}}\left(\alpha_{i}^{\mp\frac{1}{2}}\right)^*\right]\cos{\tilde
\phi} +i\left[\frac{1}{\sqrt{2}}\left(\alpha_{i}^{\pm\frac{3}{2}}\right)^*
+\frac{1}{\sqrt{6}}\left(\alpha_{i}^{\mp\frac{1}{2}}\right)^*\right]\cos{\tilde \theta} \sin\phi+\sqrt{\frac{2}{3}}\left(\alpha_{i}^{\pm\frac{1}{2}}\right)^*
\sin{\tilde \theta} \cos{\tilde \phi},
\\
A^z_{i,\pm}&=-i\left[\frac{1}{\sqrt{2}}\left(\alpha_{i}^{\pm\frac{3}{2}}\right)^*
+\frac{1}{\sqrt{6}}\left(\alpha_{i}^{\mp\frac{1}{2}}\right)^*\right]\sin{\tilde
\theta}
+\sqrt{\frac{2}{3}}\left(\alpha_{i}^{\pm\frac{1}{2}}\right)^* \cos{\tilde \theta},
\end{align}
\end{subequations}
and with Eq.~(\ref{relation}), they satisfy the relation
\begin{eqnarray}
A^j_{i,\nu}=-\nu A^{j*}_{-i,-\nu}.
\end{eqnarray}

\section{Second-order nonlinear susceptibility}
\label{appendix_nonlinear}
With a spin current of the form $\mathbb{J}=J_X\mathbf{XZ}+J_Z\mathbf{ZZ}$, where $J_X$ is the transverse amplitude, the second-order nonlinear optical susceptibility induced by the spin current is
\begin{subequations}
\begin{align}
\chi^{(2)}\left(\omega_1,\omega_2;\omega_1+\omega_2\right)
=& J_X\Big[\mathbf{XXY}\left(-2\xi_3-2\xi_3'+\xi_4'+\xi_5'\right)
+\mathbf{ZZY}\left(-4\xi_3-\xi_1'+\xi_3'-\xi_5'\right)
+\mathbf{YXX}\left(-\xi_4-\xi_5-\xi_4'-\xi_5'\right) \nonumber
\\
+ &\mathbf{XYX}\left(-2\xi_3+\xi_4+\xi_5-2\xi_3'\right)
+\mathbf{ZYZ}\left(-\xi_1+\xi_3-\xi_5-4\xi_3'\right)
+\mathbf{YZZ}\left(\xi_1-\xi_3+\xi_5+\xi_1'-\xi_3'+\xi_5'\right)
\nonumber
\\
+ &\mathbf{YYY}\left(-4\xi_3-4\xi_3'\right)\Big]
\\
+& J_Z\Big[\left(\mathbf{XYZ-YXZ}\right)\left(\xi_1+\xi_2+2\xi_3+\xi_4+3\xi_5-\xi_2'-3\xi_3'-\xi_5'\right)
\nonumber
\\
+&\left(\mathbf{YZX-XZY}\right)\left(\xi_2+3\xi_3+\xi_5-\xi_1'-\xi_2'-2\xi_3'-\xi_4'-3\xi_5'\right)
\nonumber
\\
+&\left(\mathbf{ZXY-ZYX}\right)\left(\xi_2+5\xi_3+\xi_5-\xi_2'-5\xi_3-\xi_5'\right)\Big],
\end{align}
\end{subequations}
where $\xi'_k$ is derived from $\xi_k$ by exchanging $\omega_1$ and $\omega_2$, and
\begin{subequations}
\begin{align}
\xi_1=&\left(\frac{\epsilon_r+2}{3}\right)^3|d_{\mathrm{cv}}|^2\frac{2}{3}\left[\frac{1+m_e/m_l}{(\Delta^l)^2\omega_1}
+\frac{1+m_e/m_l}{(\Delta^l)^2\Delta^l_2}-\frac{1+m_e/m_t}{(\Delta^t)^2\omega_1}
-\frac{1+m_e/m_t}{(\Delta^t)^2\Delta^t_2}\right],
\\
\xi_2=&\left(\frac{\epsilon_r+2}{3}\right)^3|d_{\mathrm{cv}}|^2\left(\frac{\Delta^l-\Delta^h}{2E_F\Delta^h\Delta^l\omega_1}+\frac{\Delta^l-\Delta^h}{2E_F\Delta^h\Delta^l\Delta^l_2}\right),
\\
\xi_3=&\left(\frac{\epsilon_r+2}{3}\right)^3|d_{\mathrm{cv}}|^2\frac{1}{5}\frac{\left(\Delta^l-\Delta^h\right)\left(\Delta^l_2-\Delta^h_2\right)}
{4E_F\Delta^h\Delta^h_2\Delta^l\Delta^l_2},
\\
\xi_4=&\left(\frac{\epsilon_r+2}{3}\right)^3|d_{\mathrm{cv}}|^2\left[\frac{\Delta^l-\Delta^h}{2E_F\Delta^h\Delta^l\omega_1}+\frac{\left(\Delta^l-\Delta^h\right)\left(\Delta^l_2+\Delta^h_2\right)}
{4E_F\Delta^h\Delta^h_2\Delta^l\Delta^l_2}\right],
\\
\xi_5=&\left(\frac{\epsilon_r+2}{3}\right)^3\frac{|d_{\mathrm{cv}}|^2}{5}\left[\frac{1+m_e/m_h}{(\Delta^h)^2\omega_1}+\frac{1+m_e/m_h}{(\Delta^h)^2\Delta^h_2}-
\frac{1+m_e/m_l}{(\Delta^l)^2\omega_1}-\frac{1+m_e/m_l}{(\Delta^l)^2\Delta^l_2}
-\frac{\Delta^l-\Delta^h}{E_F\Delta^h\Delta^l\omega_1}-\frac{\Delta^l-\Delta^h}{2E_F\Delta^h\Delta^l\Delta^l_2}
-\frac{\left(\Delta^l-\Delta^h\right)\left(\Delta^l_2+\Delta^h_2\right)}
{4E_F\Delta^h\Delta^h_2\Delta^l\Delta^l_2}\right],
\end{align}
\end{subequations}
where the factor containing the material dielectric constant
$\epsilon_r$ takes into account the difference between the macroscopic
external field and the microscopic local field.~\cite{Bloembergen}

\end{widetext}

\end{appendix}


\begin{thebibliography}{53}
\bibitem[{Not()}]{Note_on_pure_current}
\bibinfo{note}{Strictly speaking, a light beam does not conserve the
  time-reversal symmetry. But its angular momentum flux apart from the energy
  flux, being a pure spin current of interest here, does.}
\end{thebibliography}

\end{document}